\documentclass[12pt,letterpaper]{article}
\usepackage{amsfonts}
\usepackage[margin=1in]{geometry}
\usepackage{setspace}
\usepackage{natbib}
\usepackage{subfig}
\usepackage{graphicx}
\usepackage{amsmath}
\usepackage{lscape,longtable,booktabs}
\usepackage{booktabs}

\newtheorem{theorem}{Theorem}

\newtheorem{corollary}[theorem]{Corollary}

\newtheorem{definition}[theorem]{Definition}

\newtheorem{assumption}[theorem]{Assumption}
\begin{document}

\bibliographystyle{elsarticle-harv}

\title{\vspace{-1.5cm}Investing for the Long Run\thanks{The authors would like to thank Jin Su for computational support and valuable suggestions for improving the paper.}
%\\{\large\bf Preliminary and Incomplete. Do not Quote or Distribute.}
}
\author{Dietmar P.J. Leisen \\
%EndAName
{University of Mainz} \\
{Gutenberg School of Management and Economics}\\
{55099 Mainz, Germany}\\
%Phone: ++49-6131-3925542 \\
Email: leisen@uni-mainz.de
\and 
Eckhard Platen \\
%EndAName
{University of Technology Sydney} \\
{\hspace{-0.2cm}Finance Discipline Group and School of Mathematical and Physical Sciences\thanks{Also affiliated with Australian National University Canberra, Australia and University of Cape Town, South Africa}}\\
{PO Box 123, Broadway NSW 2007, Australia}\\
%Phone: ++49-6131-3925542 \\
Email: Eckhard.Platen@uts.edu.au}

\date{This version: \today}

\maketitle

\centerline{\bf Abstract} 
\quad \\[-0.5cm]
This paper studies long term investing by an investor that maximizes either expected utility from terminal wealth or from consumption. We introduce the concepts of a generalized stochastic discount factor (SDF) and of the minimum price to attain target payouts. The paper finds that the dynamics of the SDF needs to be captured and not the entire market dynamics, which simplifies significantly practical implementations of optimal portfolio strategies. We pay particular attention to the case where the SDF is equal to the inverse of the growth-optimal portfolio in the given market. Then, optimal wealth evolution is closely linked to the growth optimal portfolio. In particular, our concepts allow us to reconcile utility optimization with the practitioner approach of growth investing. We illustrate empirically that our new framework leads to improved lifetime consumption-portfolio choice and asset allocation strategies.

\quad \\[-0.3cm]
%\vspace{0.2in}

%\quad \\
\centerline{\bf Keywords}
\quad \\[-0.5cm]
stochastic discount factor, minimum pricing, optimal portfolio, growth optimal portfolio
\quad \\
\quad \\
%\quad \\
\centerline{\bf JEL Classification}
\quad \\[-0.5cm]
G11, G13

\newpage

\onehalfspacing
%\doublespacing

\section{Introduction}

Long term investing is an important concern of individual investors (e.g. saving for retirement) and of financial intermediaries (e.g. managing pension funds).  A popular advise for a moderate-risk US retirement portfolio is the 60/40 rule of thumb, i.e. to hold 60\% of marketable wealth in stocks and the remainder (40\%) in bonds (\citet{Cam&Vic:02}). However, it remains unclear when such a two-fund portfolio strategy is optimal, why these proportions are (roughly) optimal and what exactly constitutes the risky asset position. This paper aims to address such questions. In particular, we discuss practical aspects of optimal long-term investing with particular focus on identifying settings where dynamic asset allocation strategies remain rather simple.

There exists an extensive literature on the optimization of expected utility from terminal wealth and consumption-savings portfolios, including, e.g. \citet{Mer:71}, \citet{Eps&Zin:89}, \citet{Duf&Eps:92a}, \citet{Cam&Vic:99}, \citet{Cha&Vic:05}, and \citet{Kra&Sei&Ste:13}. In this literature, the martingale technique has turned out to be a promising approach; see \citet{Cox&Hua:89}, \citet{Cvi&Zap:04} and \citet{Pen:08} for details. Our paper is rooted in our observation that the martingale technique expresses desired asset allocations in terms of, so-called, stochastic discount factors (henceforth SDFs, also known as pricing kernels). 

This paper draws attention to properties of SDFs and relaxes the restrictive assumptions of classical no-arbitrage theory, which relies on the risk neutral pricing paradigm. Instead, we argue for the additional and rather natural requirement that the SDF be tradeable. We introduce and discuss the notion of minimum pricing, which allows us to use a generalized martingale technique and thereby show that our notion of tradeable SDF plays an important role in asset allocation. This addresses two of our initial questions: when are two-fund strategies optimal and what characterizes the risky asset position?

We stress the fact that the, so-called, growth optimal portfolio of the given investment universe (henceforth GP), see \citet{Kel:56} and \citet{Mer:71}, plays a central role in the solutions of most optimal investment strategies. Along these lines, minimum pricing relates to the benchmark pricing theory of \citet{Pla:02}, and \citet{Pla&Hea:10}, which opens a much wider, and thus more realistic, modeling world than available under the classical no-arbitrage paradigm. 

Our paper points out that the inverse of the GP is the natural choice of SDF and so we focus on this SDF thereafter. We draw attention to the stochastic dynamics of the GP and show that the optimal investment strategy simplifies dramatically when the dynamics of the GP is Markovian. In this, for a modeler appealing, yet also rather realistic situation, the resulting optimal strategies allocate assets into a few funds only. In particular, we show for a Markovian GP model with one driving source of uncertainty, that the optimal asset allocation consists of two funds: the riskless asset and the GP. 

Finally, we carry out an empirical evaluation of investment strategies that are derived from a particular model. Although the model does not satisfy classical no-arbitrage assumptions, its long-term dynamics are realistic and permit asset allocation in a now practically tractable manner. Our long-term investment strategies relate to higher long-term growth and achieve objectives less expensively than possible under the classical paradigm. Along these lines we also address the initial quantitative question of what constitutes an optimal proportion between riskless and risky assets. 

The literature has made tremendous advances in the theoretical understanding of asset management. It appears that many mathematical aspects have been clarified to a large extent. However, there remain intriguing problems that impede their practical relevance beyond simple situations. We contribute to this literature by addressing three formidable difficulties.

First, long term investing addresses changes in the investment opportunity set over time (\cite{Cam&Vic:02}) but solving the associated dynamic optimization problem is intrinsically hard. Moreover, the known solutions suggest that the optimal dynamic trading strategies are, in general, rather complex. Our approach, however, reduces the asset allocation to investing into a few funds, including a single risky fund that is closely related to the stochastic dynamics of the tradeable SDF.

Second, the current literature provides the impression that qualitative insights depend crucially on the underlying setup, in particular, on the concrete parametric formulation of the investment objective. For example, different preference specifications, e.g. additive utility versus recursive utility preferences, seem to lead to asset allocations that are widely different. The current paper provides a unifying theory that clarifies in optimal asset allocation the crucial link to tradeable SDFs and thereby stresses commonalities. We show that our results hold, whenever investors that maximize utility from terminal wealth or consumption aim for the least expensive strategy. 

\cite{McL&Tho&Zie:10} and \cite{Dav&Lleo:15} note that several legendary investors, including John Maynard Keynes, Warren Buffet and Bill Gross follow strategies that maximize long-term growth. Moreover, \cite{Chr:11} provides an excellent description of an older lively literature that recommends using such strategies. He explains forcefully the differences between growth maximization and utility maximization. However, the current paper employs the concept of minimum pricing to show that utility maximization in the presence of a tradeable (unknown) SDF is intrinsically linked to growth maximization. Thereby, we provide a fresh look at the older growth optimal investing literature and additional support for common long-term practitioner strategies.

Third, the practical implementation of theoretical investment strategies is currently a daunting task. Typically, it involves dynamic asset allocation in a wide range of securities to capture the impact of the theoretically possible driving factors, which represents a seemingly infeasible challenge in implementation. To make matters worse, modeling the entire market dynamics and reliably estimating, in particular, means but also covariances of a large number of assets is hard in practice; see  \cite{DeM&Gar&Upp:09}. Our approach provides a considerable simplification in this endeavor as it calls attention to modeling the dynamics of the GP. In particular, we introduce a realistic, univariate long-term dynamics for the GP, show that it can be reliably implemented, and leads to superior, two fund investment strategies.

The paper is organized as follows: Section 2 introduces our setup, while the following section discusses structural properties of price dynamics. Section \ref{sec_choice} uses these to characterize optimal portfolio choice in terms of the GP. This draws our attention to modeling the price dynamics of the GP in Section  \ref{sec_attain} which allows us to gain structural insights into portfolio construction. The following section links our results to well-known ones from the literature. Section \ref{sec_emp} evaluates the performance of our model in comparison to other approaches. Section \ref{sec_concl} concludes the paper.

\section{The Market Environment}

We assume a filtered probability space $(\Omega, \mathcal{A}, \underline{\mathcal{A}}, P)$, where $P$ denotes the real-world probability measure. $\mathcal{A}_0$ is assumed to be the trivial $\sigma$-algebra. The filtration $\underline{\mathcal{A}}=(\mathcal{A}_t)_{0 \leq t <\infty}$ describes the evolution of information in the market, i.e. at any time $t \in [0,\infty)$ the $\sigma$-algebra $\mathcal{A}_t$ describes the information available at that time. 

\subsection{Assets and Portfolios}

We consider a market composed of $d+1$ primary security accounts. These securities include the locally risk-free base line asset $S_t^0=1, t \in [0,\infty)$. 
In addition, the market is composed of $d$ risky primary security accounts $S^j=(S_t^j )_{t \in [0,\infty)}, j=1,\ldots,d$, where  all payments (e.g. dividends, income or interest) are reinvested, and $S_t^j$ is denominated in units of $S_t^0$. Note that one could also include, e.g. human capital, intellectual property, production facilities, or other (non-)financial assets in the given investment universe, as long as they can be traded. 

In line with \citet{Mer:71}, we assume perfect capital markets:  trading in the securities takes place continuously in time; there are no transaction costs; at any point in time locally risk-free, instantaneous investing and borrowing is possible; the securities are infinitely divisible, and short sales with full use of proceeds are allowed. 

For simplicity, we illustrate our approach for continuous dynamics, only, but dynamics involving jumps can be handled similarly. Also, for illustration purposes, we denominate our securities in units of the locally risk-free savings account, our baseline asset. However, our methodology allows many denominations. From a consumption-savings portfolio perspective, for example, one may prefer to denominate prices in units of the consumer-price index (CPI), i.e. employing an inflation rate accruing account as denominator. 

The dynamics of the risky $j$-th primary security account $S_t^j$ is given through the stochastic differential equation (SDE)
\begin{eqnarray}
d S_t^j & = & S_t^j \left( a_t^j  dt + \sum_{k=1}^d b_t^{j,k} d W_t^k \right) = S_t^j \left( a_t^j  dt + b_t^{j \top} d W_t \right) , \label{eq_SDE}
\end{eqnarray}
$t \in [0, \infty)$.
Here, $S_0^j>0$ is given for $j=1,\ldots,d$, and $W^1,W^2,\ldots,W^d$ are $d$ independent Brownian motions modeling the traded uncertainty. We denote by $\top$ the vector/matrix transpose and by $W_t=(W_t^1, \ldots, W_t^d)^\top$ the $d$-dimensional vector of the values of the independent Brownian motions at time $t$. We define $a_t^S=(a_t^1,\ldots,a_t^d)^\top$ as the $d$-dimensional (instantaneous) appreciation rate vector, and $b_t^S=(b_t^{j,k})_{j,k=1,\ldots,d}=(b_t^{j\top})_{j=1,\ldots,d}$ as the $d \times d$-dimensional (instantaneous) volatility matrix. Both stochastic processes are assumed to be adapted to the filtration $(\mathcal{A}_t)_{0 \leq t <\infty}$ and may be driven also by sources of uncertainty other than the traded uncertainty modeled by $W$. This allows one to capture state variables that describe some (potential) incompleteness in this market. We assume that the respective system of SDEs has a unique strong solution, see e.g. Section 7.7 in \citet{Pla&Hea:10}. 

Throughout this paper we adopt the following assumption:

\begin{assumption}\label{ass0}
Almost surely under the real-world probability measure $P$, the volatility matrix $b_t^S$ is invertible with inverse $b_t^{S,-1}$ for all $t \in [0,\infty)$, and the process $(b_t^S)_{t \in [0,\infty)}$ satisfies the condition $\int_0^T \sum_{j,k=1}^d (b_t^{j,k})^2 dt < \infty$ for all $T \in [0, \infty)$.
\end{assumption}

The invertibility assumption for the volatility matrix ensures that the $d$-dimensional vector 
\begin{equation}
\theta_t = b_t^{S,-1} a_t^S  \label{eq_mpr}
\end{equation}
is well-defined at all times, and we refer to $\theta_t$ as the market-price-of-risk vector at time $t$. A consequence is that $\theta_t$ becomes the volatility of the unique and finite growth optimal portfolio (GP), as we discuss later. The second condition in Assumption \ref{ass0} ensures sufficient integrability of the stochastic It\^{o} integrals involved. 

\subsection{Preferences}\label{sec_PS_subsec_hedge}

We introduce an indicator variable $\chi \in \{0,1 \}$, fix $T \in [0,\infty)$ and denote by $V_0$ the investor's initial wealth, as well as, by $\pi=(\pi_t)_{0 \leq t \leq T}$ the (vector) process of wealth weights, with $\pi_t=(\pi_t^1, \pi_t^2, \ldots, \pi_t^d)^\top$. Here, $\pi_t^j$ denotes the relative weight of the time $t \in[0,T]$ investment in the risky primary security $S_t^j$, $j=1,\ldots,d$. Note that $\pi_t^0=1-\sum_{j=1}^d \pi_t^j$ is the fraction invested in the locally riskfree baseline asset. (Negative weights indicate borrowing.) 
We denote by $C=(C_t)_{0 \leq t \leq T}$ the agent's consumption process, financed by the initial capital $V_0$ and through the investment strategy $\pi$. Then, the agent's wealth after consumption is given by the process $(V_t^\pi)_{0 \leq t \leq T}$, satisfying by \eqref{eq_mpr} the SDE
\begin{eqnarray}
dV_t^\pi & = & \left( \pi_t^\top a_t^S V_t^\pi - \chi C_t \right) dt + (\pi_t^\top b_t^S) V_t^\pi d W_t  \label{eq_wealth} \\
& = & \left( \pi_t^\top b_t^S \theta_t V_t^\pi - \chi C_t \right) dt + (\pi_t^\top b_t^S) V_t^\pi d W_t  . \label{eq_wealth2}
\end{eqnarray}
%The wealth process is fully characterized by the choice of the weight process $\pi$ and the consumption process $C$ (in the consumption-savings portfolio choice problem; $\chi=1$) and $\chi C=0$ (in the asset allocation problem; $\chi=0$).

For a given parameter $\varepsilon \geq 0$ and given indicator variable $\chi \in \{0,1\}$, our objective is to determine a process $J=(J_t)_{t  \in [0,T]}$ that solves the optimization
\begin{equation}
J_t=\max_{\pi,C} E\left[ \left. \int_t^T \chi f(C_s,J_s,s) ds + \varepsilon B (V_T^\pi) \right| \mathcal{A}_t  \right] , \label{setup_opt}
\end{equation}
for all $t \in [0,T)$. 
Here, at any time $t \in [0,T)$, the maximization is taken over the remaining time period $[t,T]$ by choosing a respective portfolio weight process $\pi$ and consumption process $C$ with associated wealth process \eqref{eq_wealth2}. Furthermore, we employ in \eqref{setup_opt} the conditional expectation $E[\cdot | \mathcal{A}_t ]$ under the real-world probability measure $P$, given the information at time $t$, encapsulated in $\mathcal{A}_t$. The time $t$ information includes the value of the portfolio $V_t^\pi$ at time $t$. The process $J$ characterizes the, so-called, life-time utility derived from consumption.

Throughout this paper we adopt, for simplicity, the following additional assumption on the utility functions involved:

\begin{assumption}\label{ass2}
For given $s \in[0,T]$ and $l \in \mathbb{R}$ the functions $f(\cdot,l,s),B(\cdot):\mathbb{R}^+ \to \mathbb{R}$ are twice differentiable, strictly increasing, strictly concave and fulfill the Inada conditions, i.e. $\lim_{x \downarrow 0} f' (x,l,s)=\lim_{x \downarrow 0} B'(x)=\infty$ and $\lim_{c \to \infty} f' (c,l,s)=\lim_{x \to \infty} B'(x)=0$.
\end{assumption}

%\begin{assumption}\label{ass2a}
%In consumption-savings problems we assume that $f$ is uniformly Lipschitz, i.e. there exists a constant $K>0$ such that a.s. $|f(c,l,t) - f (\tilde{c}, \tilde{l},t)| \leq K (|c-\tilde{c}|+|l-\tilde{l}|)$ for all $c,\tilde{c} \in \mathbb{R}^+; l,\tilde{l} \in \mathbb{R}$.
%\end{assumption}

The parameters $\varepsilon$ and $\chi$ allow us to study a variety of portfolio choice problems. 
The case $\chi=0$, $\varepsilon >0$ corresponds to the problem of a price taking agent who maximizes expected utility derived from terminal wealth $V_T^\pi$, as in  \citet{Mer:71}, i.e. a portfolio choice problem without consumption (a so-called asset allocation problem); the function $B$ plays here the role of the utility function. 

The case $\chi=1$ corresponds to a consumption-savings problem (or intertemporal consumption and portfolio choice problem) either with bequest ($\varepsilon>0$) or without bequest ($\varepsilon=0$). 
The parameter $\varepsilon$ controls the  importance of bequest relative to the utility derived from consumption. 
Our specification in \eqref{setup_opt} captures, among others, the usual case of time-separable utility, as well as, the continuous-time version of, so-called, Epstein-Zin preferences introduced within the context of stochastic differential utility by \citet{Duf&Eps:92}, see Appendix \ref{appA}. We will discuss in Subsections \ref{subsec_pref1} and \ref{subsec_pref0} examples of time-additive preferences over consumption and of preferences over terminal that are covered. Whenever needed, we assume existence and uniqueness of stochastic differential utility in the sense of \citet{Duf&Eps:92}. 

In a general market setting, the conditional expectation \eqref{setup_opt} is usually a rather complex function of many state variables. This makes practical applications of the theoretical optimal strategy difficult, if not seemingly impossible: (too) many quantities have to be accurately modeled and estimated to obtain a practically useful strategy. It is well known that the estimation of parameters characterizing expected returns requires longer observation windows than available in reality, see e.g. \citet{Kan&Zho:07} and \citet{DeM&Gar&Upp:09}. Furthermore, estimated covariance matrices of returns are stochastic themselves and, in general, extremely difficult both to estimate and to invert with meaningful outcomes, especially for a large number of assets, see e.g. \citet{Bai&Ng:02}, \citet{Lud&Ng:07} and \citet{Okh&Scm:06}. Notwithstanding these difficulties in general, the main emphasis of this paper is to point out realistic and practically relevant situations, where the conditional expectation in \eqref{setup_opt} becomes a function of the current time $t$ and potentially a few, well observable state variables. 

\section{Structural Properties of Price Dynamics}\label{sec3}

A crucial determinant of asset allocation is the price dynamics of securities. To prepare for our asset allocation analysis in the next section we, therefore, characterize some of their structural properties. The first subsection discusses pricing through a (generalized) stochastic discount factor (SDF). The second subsection stresses the tradeability of the SDF. The following subsection identifies tradeable SDFs, i.e. they are given as the inverse of the, so-called, growth optimal portfolio (GP). The fourth subsection discusses the benchmark pricing theory and introduces minimum pricing. The fifth subsection explains how minimum pricing is useful for analyzing optimal portfolio allocations.

\subsection{Stochastic Discount Factor}\label{sec3_subsec1}

The absence of arbitrage opportunities in market dynamics is a common assumption in classical financial economics, often employed to ensure well-defined financial markets. The workhorse underpinning modeling efforts is the Fundamental Theorem of Asset Pricing, which asserts that the absence of, so-called, classical arbitrage opportunities is equivalent to the existence of an equivalent risk-neutral probability measure; see \citet{Har&Kre:79} and \citet{Del&Sch:06}. In the language of financial economics this is equivalent to the existence of a stochastic discount factor (SDF) with respective properties; see e.g. \citet{Coc:01}. 
Unfortunately, staying in the classical no-arbitrage framework comes at the cost of restrictive and, potentially, unrealistic modeling assumptions that impede its application in practice; see e.g. \citet{Loe&Wil:00} and \citet{Bal&Gra&Pla:15}. 

This paper takes a more general approach, in line with a growing literature that avoids classical no-arbitrage assumptions and recommends working in a wider modeling world than the one that assumes the existence of an equivalent risk-neutral probability measure; see e.g. \citet{Loe&Wil:00}, \citet{Pla:02}, \citet{Pla:06}, \citet{Kar&Kar:07}, \citet{Fer&Kar:10} and \citet{Pla&Hea:10}. Compared to the more general approach we pursue, the consequences of the classical, more restrictive assumptions for short term investing may be minor. However, the long-term consequences are substantial, as we demonstrate.

By generalizing concepts and ideas presented in the wide literature on classical asset pricing that have been well summarized in \citet{Coc:01}, let us introduce the following assumption: 

\begin{assumption}\label{ass1}
There exists an adapted, strictly positive process $(F_s)_{0 \leq s \leq T}, T \in [0,\infty)$ such that $0 < F_t < \infty$ almost surely for $t \in [0,T]$, and such that for all primary securities $S_t^j$, $j=0,\ldots,d$, and value processes of self-financing portfolios the SDE for the product of $F_t$ and the value or price process is driftless, in particular, $(F_t \cdot S^j_t)_{0 \leq t \leq T}$ is driftless. To be even more precise, we assume that we can write
\begin{equation}
d F_t = F_t ( a_t^F dt + b_t^{F \top} dW_t + c_t^{F \top} d \bar{W}_t)  \label{eq_SDF}
\end{equation}
with $F_0=1$ for a suitable adapted real valued process $a^F$, a suitable adapted vector process $b^F=(b^{F1}, \ldots, b^{Fd})^\top$, and a suitable adapted vector process $c^F=(c^{F1}, \ldots, c^{Fn})^\top$, with non-negative integer $n$. Here, the non-traded uncertainty $\bar{W}=(\bar{W}^1, \ldots, \bar{W}^n)^\top$ is given by a vector process of $n$ independent Brownian motions that are independent of the Brownian motion vector process $W$, the traded uncertainty. We refer to $F$ with these properties as a \emph{stochastic discount factor (SDF)}. 
\end{assumption}

This assumption is weaker than the classical one, where the product process $(F_t \cdot S_t^j)_{0 \leq t \leq T}$ must form an $(\underline{\mathcal{A}},P)$-martingale. Here we only require these processes to form local martingales. Assumption \ref{ass1} allows us to discuss in a more flexible, transparent and unified manner different trade-offs in terms of pricing and asset allocation that a modeler faces. In particular, the well-established literature on asset pricing relates conveniently, as a special case, to our approach.

The common modeling approach starts with assuming the existence of a risk-neutral probability measure. Such a restriction blends together two conditions that we both relax in Assumption \ref{ass1}. First, under classical no-arbitrage assumptions, one typically assumes that for any price process $(A_t)_{t\in[0,T]}$ the product $(A_t F_t)_{0 \leq t \leq T}$ forms a true martingale, which leads to the pricing rule
\begin{equation}
A_t F_t = E [A_s F_s | \mathcal{A}_t]  \label{eq_EMM_pricing}
\end{equation}
for all $0 \leq t \leq s \leq T$. In Assumption \ref{ass1} we only request that $A_t F_t$ is driftless, and, thus, forms a local martingale. Second, the pricing rule \eqref{eq_EMM_pricing} recovers so-called classical risk-neutral pricing, see \citet{Coc:01}, \emph{if and only if} $(F_t)_{0 \leq t \leq T}$ forms a true martingale. The risk-neutral probability measure is in this case characterized via the Radon-Nikodym density process $(F_t)_{0 \leq t \leq T}$. We underline the crucial fact that Assumption \ref{ass1} permits a much wider class of models where $F_t$ needs only to form an $(\underline{\mathcal{A}},P)$-local martingale, and the pricing rule \eqref{eq_EMM_pricing} is only one of many possible ones.

As we will see later on, it is a rather strong assumption of the common classical modeling approach that an asset price multiplied by an SDF needs to form a true martingale. We emphasize once more that we do \emph{not} assume this property in the current paper. Consequently, the pricing rule \eqref{eq_EMM_pricing} does not need always to hold in our more general setting. Furthermore, even if the pricing rule would hold, this may not necessarily imply that $(F_t)_{0 \leq t \leq T}$ itself needs to form a true martingale. This does not mean that we could not have in our market price processes that result from formally applying the risk-neutral pricing rule. However, we only request in Assumption \ref{ass1} that the resulting price processes, when multiplied with the SDF, have no drift, thus form local martingales.

While the local martingale property in Assumption \ref{ass1} is significantly relaxed (compared to the martingale property), it imposes still restrictions. These allow us to derive the following conclusions: First, Assumption \ref{ass1} in \eqref{eq_SDF} implies with respect to the locally risk-free asset $S_t^0=1$ that $a_t^F=0$. Furthermore, by equations \eqref{eq_SDE} and \eqref{eq_mpr}, it follows via the It\^{o} formula with respect to all risky securities $S_t^j$ with $j=1,\ldots,d$  that
\begin{eqnarray*}
d (F_t S_t^j) & = & F_t a_t^j S_t^j dt + F_t b_t^{j \top} S_t^j dW_t + S_t^j F_t b_t^{F \top} dW_t + S_t^j F_t c_t^{F \top} d \bar{W}_t + b_t^{j \top} b_t^F F_t S_t^j dt \\
& = & S_t^j F_t  \left( a_t^j + b_t^{j \top} b_t^F \right) dt + S_t^j F_t \left(b_t^{j \top} + b_t^{F\top} \right) dW_t  + S_t^j F_t c_t^{F \top} d \bar{W}_t \\
& = & S_t^j F_t \left\{ b_t^{j \top} \left( \theta_t  + b_t^{F} \right) dt + \left(b_t^{j \top} + b_t^{F\top} \right) dW_t + c_t^{F \top} d \bar{W}_t \right\} .
\end{eqnarray*}
For this SDE to be driftless, we must have $b_t^{j \top} \left( \theta_t  + b_t^{F} \right)=0$ for all $j=1,\ldots,d$. This means that we need to have $b_t^{F} = - \theta_t$, i.e.  by \eqref{eq_SDF} we obtain
\begin{equation}
\frac{d F_t}{F_t}  = - \theta_t^\top dW_t + c_t^{F \top} d \bar{W}_t .   \label{eq_SDF2}
\end{equation}
%or equivalently
%\begin{equation}
%d \left(\frac{1}{F_t} \right)= \frac{1}{F_t} \left\{ \theta_t^\top \theta_t dt + \theta_t^\top dW_t + c_t^{F \top} c^F dt -  c_t^{F \top} d \bar{W}_t \right\} . \label{eq_inv_SDF}
%\end{equation}
%
%Equation \eqref{eq_SDF2} provides us with an SDF $(F_t)_{0 \leq t \leq T}$ that makes the product with all primary security price processes driftless (a local martingale). In fact, a stronger observation can be made directly from here: For any self-financing portfolio price process $(A_t)_{0 \leq t \leq T}$ we know then that its price process multiplied with the SDF, that is  $(A_t F_t)_{0 \leq t \leq T}$, forms a \emph{local} martingale. I THINK THIS REQUIRES A SHORT CALCULATION. LATER WE DO SUCH A CALCULATION AND DERIVATION. BUT HERE WE DO NOT NEED THIS AND SO I TEND TO CUT THE LAST SENTENCE OR THE ENTIRE PARAGRAPH.
We emphasize that this is the generic form of an SDF in our market.

\subsection{Tradeability of SDF}

Our definition of an SDF in Assumption \ref{ass1} explicitly allows the SDF to be driven by traded and potentially also by non-traded uncertainty, see equation \eqref{eq_SDF}. Therein, the SDF $F=(F_t)_{0 \leq t \leq T}$, satisfying equation \eqref{eq_SDF2} can be represented as
\begin{equation}
F_t = F_0 \exp \left\{ - \frac{1}{2} \int_0^t \theta_s^{\top} \theta_s ds - \int_0^t \theta_s^{\top} d W_s \right\} \tilde{F}_t ,
\label{eq_Ftilde_help3}
\end{equation}
with factor
\begin{equation}
\tilde{F}_t = \exp \left\{ - \frac{1}{2} \int_0^t c_s^{F \top} c_s^F ds + \int_0^t c_s^{F \top} d\bar{W}_s \right\} . \label{eq_Ftilde_help}
\end{equation}
The process $\tilde{F}$ is driven by non-traded uncertainty and, therefore, cannot be hedged through dynamic asset allocation. The decomposition \eqref{eq_Ftilde_help3} shows that introducing non-tradeable uncertainty into the SDF is equivalent to introducing a non-hedgeable factor process $\tilde{F}$ in the construction of the SDF. Theoretically, this factor process seems to provide wider generality for the notion of an SDF. However, as we will argue now, it appears not to provide any usable flexibility in practical valuations or hedging. 

The factor $\tilde{F}$ is not unique but, by definition, any choice of $\tilde{F}$ must not be driven by any traded uncertainty. Taking a formal look at this fact from the valuation perspective, one notes that, in order to make the pricing rule \eqref{eq_EMM_pricing} work, say, under classical assumptions, the process $\tilde{F}$ has to be a true martingale. As a consequence of basic rules of stochastic calculus, the non-traded uncertainty $\tilde{F}$ in \eqref{eq_Ftilde_help3} does not impact in \eqref{eq_EMM_pricing} the prices of any hedgeable payoffs and is, therefore, redundant in such a valuation approach. As such it is superfluous when pricing replicable claims.

There is an additional conceptual argument against introducing a superfluous factor $\tilde{F}$ into an SDF, taken from the perspective of practical implementation of optimal portfolio choices. We will see in the next sections that optimal portfolios and their price processes become driven by the SDF. Therefore, when the factor $\tilde{F}$ is not replicable from traded security prices, i.e. when non-traded sources of uncertainty drive the SDF, then optimal portfolios and their price processes become driven by non-tradeable uncertainty, i.e. require investing in non-tradeable factors. To implement such strategies, we would have to make non-tradeable uncertainty tradeable, which defeats the purpose. 

In summary, the generality of the forms of the dynamics of the SDF satisfying \eqref{eq_SDF2}, which involves the factor $\tilde{F}$, is artificial and evaporates from the perspective of practical feasibility. 
Based on these insights, in the remainder of this paper we remove from the SDF, given in \eqref{eq_Ftilde_help3}, the superfluous non-tradeable factor $\tilde{F}$. Instead we offer more practically relevant flexibility in the current paper than under classical assumptions: we allow $(A_t F_t)_{t \in[0,T]}$ to form strict local martingales (our approach) instead of true martingales (classical approach). We will demonstrate that this flexibility is important for making long-term investing less expensive.

Consequently, throughout this paper, we work with an SDF satisfying \eqref{eq_SDF2} with $c_t^F=0$, that is the SDF satisfies from now on the SDE
\begin{equation}
d F_t = - F_t \theta_t^\top d W_t , \label{eq_SDF3}
\end{equation}
for $t \geq 0$ and $F_0=1$.

%Our SDF definition leads us beyond classical assumptions in finance. As a consequence, our general setting provides access to modeling and exploiting new phenomena, including the possibility of designing less expensive portfolios for hedging contingent claims and optimizing expected utilities, as we will illustrate later on.

\subsection{The Inverse Growth Optimal Portfolio as SDF}\label{rw_price}

The previous subsection introduced the notion of a tradeable SDF, which we identify in this subsection and for which we discuss here  implications for pricing.

The growth optimal portfolio (GP) with value $V_t^{\pi^{GP}}=V_t^{GP}$ at all times $t \in[0,T]$, characterized as the portfolio with $V_0^{GP}=1$, is unique and has wealth weights 
\begin{equation}
\pi_t^{GP}=b_t^{S,-1,\top} \theta_t \label{eq_wealth_weight_GP}
\end{equation}
at time $t \in[0,T]$ (without consumption, $C=0$). Based on equation \eqref{eq_wealth2} it fulfills the  SDE
\begin{eqnarray}
d V_t^{GP} & = & ( \pi_t^{GP \top}  a_t^S ) V_t^{GP} dt + \pi_t^{GP\top} b_t^S V_t^{GP} d W_t , \nonumber \\
& = & V_t^{GP} \left\{  (\theta_t^{\top} \theta_t ) dt + \theta_t^{\top} d W_t \right\}  \label{eq_dyn_GP} 
\end{eqnarray} 
for $t \geq 0$ with $V_0^{GP}=1$. By Assumption \ref{ass0} the GP exists in our market, i.e. due to its finite expected instantaneous growth rate $\frac{1}{2} \theta_t^\top \theta_t$, it is strictly positive and finite at any finite time. By application of the It\^{o} formula to $1/V_t^{GP}$ we find that
\begin{equation}
F_t = \frac{1}{V_t^{GP}}  \label{eq_Ftilde_help2}
\end{equation}
for $t \geq 0$ solves the SDE \eqref{eq_SDF3}. Thus, the inverse of the GP is a suitable choice of an SDF in the sense of Assumption \ref{ass1}. 

Additionally, the SDF \eqref{eq_Ftilde_help2} is tradeable. By Assumption \ref{ass1}, any other candidate for a tradeable portfolio that can be used as a num\'{e}raire should, when used as denominator for the GP, generate a strictly positive local martingale, which by Fatou's Lemma is then a supermartingale, see \citet{Pla&Hea:10} Theorem 10.2.1. But also the GP, when used as denominator for the candidate portfolio, is a local martingale, and, thus, a supermartingale. By Jensen's inequality, the candidate portfolio expressed in units of the GP can only equal the constant one. Therefore, the tradeable SDF \eqref{eq_Ftilde_help2} is \emph{unique}.

In the remainder of this paper we set $F$ according to equation \eqref{eq_Ftilde_help2}, or equivalently according to the SDE \eqref{eq_SDF3}. 

\subsection{Benchmark Pricing Theory}

The benchmark pricing theory, proposed by \citet{Pla:02}, assumes only the existence of the GP, which it calls benchmark, and makes it its central building block. 
Any value multiplied with $F_t$ is called a benchmarked value. In particular, we denote by $\hat{S}_t^{j}=\frac{S_t^j}{V_t^{GP}}$ the respective benchmarked value of the primary risky security $S_t^j$, $j=1\ldots,n$, and consider the \emph{benchmarked wealth} process $\hat{V}^\pi = \frac{V_t}{V_t^{GP}}$ without consumption ($C=0$). (The next subsection will study \emph{benchmarked consumption-adjusted wealth} processes $\hat{G}^\pi$.)

Through the existence of the GP, see \citet{Pla&Hea:10}, there is no economically meaningful arbitrage in the market, in the sense that no strategy can generate in finite time a strictly positive portfolio with infinite wealth from finite initial capital. Under classical no-arbitrage assumptions, \citet{Lon:90} made the observation that the risk-neutral price can be recovered by choosing the GP as num\'{e}raire and the real world probability measure as pricing measure. This means, the pricing rule \eqref{eq_EMM_pricing} with $F_t = 1/V_t^{GP}$ recovers in the classical setting the risk-neutral price for a replicable contingent claim. Therefore, the GP is also termed the num\'{e}raire portfolio (NP).

The benchmark pricing theory goes much further than reformulating classical risk-neutral pricing, it drops the classical assumptions, see \citet{Pla&Hea:10}, and requests instead only the existence of the GP. It employs then the GP as denominator, num\'{e}raire and benchmark. In summary, the benchmark pricing theory obtains a unique and \emph{tradeable} SDF by setting $F_t$ equal to the locally riskless baseline security, denominated in units of the GP. This means that the SDF $F_t$ represents the inverse of the discounted GP. Therefore, by setting $\tilde{F}_t=1, t \geq 0$ in \eqref{eq_Ftilde_help}, we identify via \eqref{eq_Ftilde_help2} the inverse $F_t^{-1}$ with the discounted benchmark $V_t^{GP}$. 

In the general setting of the benchmark pricing theory, several self-financing portfolios can replicate the same payoff. While this means that the classical Law of One Price may not hold any longer in the more general modeling world we consider, this does not permit creation of any economically meaningful arbitrage: the absence of economically meaningful arbitrage should be interpreted in the sense that there does not exist a strictly positive portfolio that has an infinite instantaneous growth rate such that it could become infinite in finite time; see \citet{Pla&Hea:10}. Most importantly, from a conceptual viewpoint we note that the failure of the Law of One Price allows several pricing rules to coexist. For instance, one can formally apply the risk-neutral pricing rule even when $F$ does not form a true martingale. However, as we will see now, there exists always a minimum price, which is uniquely determined. 

As indicated earlier, the growth optimal portfolio (GP) maximizes expected logarithmic utility derived from terminal wealth. 
The GP is the best performing, strictly positive portfolio in the sense that in the long run its value surpasses almost surely the value of all other strictly positive portfolios, see e.g. \citet{Pla&Hea:10}, Theorem 10.5.1. This also means that the inverse of the GP, our SDF, is the inverse of the portfolio that leads, in the long run, almost surely to lower values than the inverse of any other strictly positive portfolio could achieve. When we employ the inverse of the GP as SDF we, therefore, can  intuitively expect it to lead us to the lowest prices possible among all potential price processes for a targeted payoff. This means, when one employs the benchmarked savings account as SDF, one obtains intuitively via the pricing rule \eqref{eq_EMM_pricing} the minimal possible price process. 

To make this precise we introduce the notion of a fair price or value process:

\begin{definition}\label{defn6}
A price or value process $(A_t)_{0 \leq t \leq T}$ is called fair, when it satisfies  the pricing rule \eqref{eq_EMM_pricing} with the inverse of the GP as SDF $F=1/V^{GP}$, that is
\begin{equation}
\frac{A_t}{V_t^{GP}} = E \left[ \left. \frac{A_s}{V_s^{GP}} \right| \mathcal{A}_t \right] \label{eq_RW_pricing}
\end{equation}
for $0 \leq t \leq s \leq T$. 
\end{definition}

This means that the benchmarked value $\hat{A}_t = \frac{A_t}{V_t^{GP}}=F_t A_t$ forms an $(\underline{\mathcal A},P)$-martingale. In formula \eqref{eq_RW_pricing} pricing is performed under the real-world or objective probability measure $P$ with the inverse of the GP as SDF. In \citet{Pla&Hea:10} this is referred to as \emph{real-world pricing} and the formula \eqref{eq_RW_pricing} is called the real world pricing formula.

Assumption \ref{ass1} requires that all benchmarked prices form local martingales. This implies that all benchmarked wealth processes (without consumption) are local martingales and, thus, supermartingales. It is known, see Lemma A1 in \citet{Du&Pla:16}, that a martingale is the minimal non-negative supermartingale that delivers at a bounded stopping time a targeted nonnegative integrable payout. This fact implies that the \emph{benchmarked} wealth process, which aims for the minimal possible initial expense to deliver some specified payoff, must be a martingale, and, thus, by Definition \ref{defn6} it must be a fair price process. The respective martingale exists and is unique because its benchmarked value at a given time is the conditional expectation of the benchmarked portfolio value at maturity under the available information. 

We summarize this insight as follows:

\begin{corollary}\label{newcor6}
The minimal price or value process that delivers a targeted payoff needs to be fair. More expensive price or value processes are possible that when benchmarked form non-negative local martingales, and, thus, supermartingales.
\end{corollary}

It is important to emphasize that our assumptions assure the existence of the GP, and avoid the restrictive classical no-arbitrage assumptions. In particular, we avoid the request on the existence of an equivalent risk-neutral probability measure. This opens up a much wider modeling world than provided under classical no-arbitrage assumptions. 
%In particular, long-term dynamics and phenomena can be more realistically handled, as we will illustrate later on. Economically meaningful arbitrage is excluded since the portfolio that performs best in the long run, the GP, remains finite for finite time.

\subsection{Minimum Pricing of Consumption-Savings Investments}\label{subsec_minprice}

The previous subsection introduced the notion of real-world pricing and showed that it leads to minimal possible prices. Let us now apply this pricing approach for consumption-savings investments. 

Recall that we denote for any traded security with value $S_t^j$ by $\hat{S}_t^{j}= \frac{S_t^j}{V_t^{GP}} = S_t^j F_t$ its benchmarked  value. For a given consumption-savings investment strategy $(\pi,C)$ with wealth process defined through equations \eqref{eq_wealth}-\eqref{eq_wealth2} and benchmarked wealth process $\hat{V}_t^\pi =V_t F_t$, we introduce for $t \in [0,T]$ the benchmarked \emph{consumption-adjusted} wealth process $\hat{G}^\pi$ by setting
\begin{equation}
\hat{G}_t^\pi=\hat{V}_t^\pi + \chi \int_0^t \hat{C}_s ds ,
\label{eq_Ftilde_help4}
\end{equation}
with benchmarked consumption
\begin{equation}
\hat{C}_t^\pi = C_t^\pi F_t .
\label{eq_Ftilde_help5}
\end{equation}

Note in \eqref{eq_Ftilde_help4} that the accumulated benchmarked consumption is added to benchmarked wealth to give benchmarked consumption-adjusted wealth. This is different to the methodology in most papers on consumption-savings problems, where accumulated consumption is added to wealth to give consumption-adjusted wealth. Our choice in \eqref{eq_Ftilde_help4} is important for obtaining conveniently the least expensive strategy that produces a targeted payoff stream.

Equations \eqref{eq_wealth2} and \eqref{eq_SDF3} imply for the benchmarked wealth $\hat{V}_t^\pi$ the SDE
\begin{eqnarray}
d \hat{V}_t^\pi & = & F_t d V_t^\pi  - V_t^\pi F_t \theta_t^\top d W_t - (\pi_t^\top b_t^S)  V_t^\pi F_t \theta_t dt \nonumber \\
& = & (\pi_t^\top b_t^S  - \theta_t ) \hat{V}_t^\pi d W_t - \chi \hat{C}_t dt , \label{eq_Ftilde_help7}
\end{eqnarray}
which yields by \eqref{eq_Ftilde_help4}:
\begin{equation}
d \hat{G}_t^\pi = \hat{V}_t^\pi  (\pi_t^\top b_t^S  - \theta_t ) d W_t  . \label{eq_Ftilde_help8}
\end{equation}
This means that the benchmarked self-financing portfolio process $\hat{G}^\pi$ is driftless and, therefore, forms a local martingale. In addition, $\hat{G}_t^\pi$ is nonnegative. Thus, again by Fatou's Lemma, $\hat{G}^\pi$ is a supermartingale.

Note that benchmarked (consumption-adjusted) wealth processes may exist in our setting that are strict supermartingales but not martingales. This means that these portfolios may not be fair. However, it is interesting to further study those \emph{particular} portfolios which \emph{are} fair.

%\begin{definition}\label{defn1}
%A consumption-savings investment strategy $(\pi,C)$ is called \emph{fair} if its weighted (consumption-adjusted) value process is a martingale, i.e. for $0 \leq s \leq t \leq T$:
%\begin{equation}
%\hat{G}_s^\pi = E[\hat{G}_t^\pi | \mathcal{A}_s] . \label{eq_Ftilde_help6}
%\end{equation} 
%\end{definition}

Due to the classical Law of One Price, usually, the literature does not emphasize that the investor cares about the minimal possible initial expense. Nevertheless, in our more general setting, where several self-financing portfolios may exist that deliver the same payout,  this objective now appears as a reasonable component of the optimization. Lower initial capital needed for a payout translates into extra capital that allows for higher consumption. Therefore, we aim for consumption-savings profiles and associated wealth processes that require minimal initial capital. 

In the remainder we consider the optimization problems where the investor restricts herself to fair consumption-savings portfolios in the sense of Definition \ref{defn6}, such that her objective is to determine a process $J=(J_t)_{t  \in [0,T]}$ that solves, as in \eqref{setup_opt}, the optimization
\begin{equation}
J_t=\max_{(\pi,C)} E\left[ \left. \int_t^T \chi f(C_s,J_s,s) ds + \varepsilon B (V_T^\pi) \right| \mathcal{A}_t  \right] , \label{setup_opt2}
\end{equation}
for all $t \in [0,T)$, now subject to \emph{fair} and self-financing wealth dynamics \eqref{eq_wealth2}. 

The utility optimization problem \eqref{setup_opt2} may be different from our initial utility optimization \eqref{setup_opt}, since cost minimization is not an explicit goal in \eqref{setup_opt} and so the investor may seem to have opportunities to trade off its benefits against potential utility gains. Whenever needed, we assume existence and uniqueness of stochastic differential utility in the sense of \citet{Duf&Eps:92} over this restricted set of strategies. 

Along the way, significant practical benefits become available. As we will see in the next sections, the restriction to fair strategies permits us to proceed \emph{analogous} to the well-known martingale technique. Intuitively, the optimization problem \eqref{setup_opt2} means that the investor carries out a two-step optimization along the lines of the martingale technique: First, she looks for a fair optimal consumption profile for any given initial cost and then, in the second step, ensures that the initial cost matches her initial wealth.

Under classical no-arbitrage assumptions, the pricing rule \eqref{eq_EMM_pricing} is the risk-neutral pricing rule and has been the prevailing one. As we will see later on, by applying formally risk-neutral pricing to our more general setting (which is possible) one obtains a benchmarked portfolio that is a local martingale. However, it may not be a martingale. In any case, it would be a supermartingale and, when not a martingale, would be more expensive than the respective martingale, the fair price. Real-world pricing according to \eqref{eq_RW_pricing} is minimum pricing and yields fair consumption-adjusted wealth processes. The restrictive assumption on the existence of an equivalent risk-neutral probability measure is, in our general setting, no longer enforced and portfolios can be constructed with less expensive dynamic asset allocation strategies than those obtained from formally applied risk-neutral pricing. 

Minimum pricing (real-world pricing) is to be distinguished from risk-neutral pricing, where a risk-neutral probability measure $Q$ is assumed to exist in a probability space $(\Omega, \mathcal{A}, \underline{\mathcal{A}}, Q)$ s.t. savings account discounted price processes become martingales under the assumed risk-neutral probability measure. Under the classical no-arbitrage paradigm it is well-known that an SDF can be used to define the risk-neutral probability measure, see \citet{Coc:01}. Risk-neutral and real-world pricing are equivalent under the existence of an equivalent risk-neutral probability measure, see equation (10.4.5) in \citet{Pla&Hea:10}. However, they differ in theory and in actual markets, see e.g. \citet{Bal&Gra&Pla:15}.

By formally applying risk-neutral pricing, which is currently mostly done in practice, one simply ignores the possibility that the benchmarked savings account may be, in reality, a strict local martingale, and, thus, a strict supermartingale. In such a case, the formally obtained risk-neutral price would be usually more expensive than the minimum price which is obtained via the real-world pricing formula \eqref{eq_RW_pricing}.

An important question that arises in our optimization is the tradeability of the fair optimal consumption-savings portfolio. The next sections will show that the fair portfolio is tradeable once the SDF is tradeable. In summary, \emph{fair} portfolio strategies encapsulate already an important first optimization step that does not arise under classical assumptions, where one formally applies risk-neutral pricing.

\section{Optimal Benchmarked Portfolio Choice}\label{sec_choice}

The previous sections discuss the central role of the GP as benchmark or num\'{e}raire in pricing.
This section assumes real-world pricing to obtain minimum prices and uses the properties derived to re-express the optimal wealth process (as well as the optimal consumption process) in terms of the GP and the baseline security. 

Let us consider a budget-feasible, fair consumption-savings and terminal wealth plan and postpone to a further discussion the issue, whether such a plan would be tradeable. The problem then reduces to finding the optimal, budget-feasible plan, i.e. we look for processes $C$ and $J$, and a random variable $V_T^\pi$ that maximize the right-hand side in equation \eqref{setup_opt2}. 

Real-world pricing tells us that the current benchmarked value of a consumption-savings investment strategy $(\pi,C)$ with benchmarked consumption-adjusted value process  $\hat{G}^\pi$ is given by the conditional expectation
\begin{eqnarray*}
\hat{G}_t^\pi & = & E \left[ \left. \hat{G}_T^\pi \right| \mathcal{A}_t \right] = E \left[ \left. \int_0^T \chi \frac{C_s}{V_s^{GP}} ds + \frac{V_T^\pi}{V_T^{GP}} \right| \mathcal{A}_t \right]  .
\end{eqnarray*}
Therefore, we have for the benchmarked portfolio value $\hat{V}_t^\pi=\frac{V_t^\pi}{V_t^{GP}}$ and the benchmarked consumption $\hat{C}_t=\frac{C_t}{V_t^{GP}}$ the equality
\begin{eqnarray}
\hat{V}_t^\pi & = & E \left[ \left. \int_t^T \chi \hat{C}_s  ds + \hat{V}_T^\pi \right| \mathcal{A}_t \right] ,   \label{eq_value}
\end{eqnarray}
$t \in [0,T]$. Using a Lagrange multiplier $\lambda$ to capture the initial budget constraint, we have by \eqref{setup_opt2} to maximize at the initial time $t=0$ the following expression:
\begin{eqnarray}
& & \max_{\pi,C} E\left[ \int_0^T \chi f(C_s,J_s,s) ds + \varepsilon B (V_T^\pi) \right] - \lambda \left( E \left[ \hat{G}_T^\pi \right] - \hat{G}_0 \right)  \label{eq_Lagrange1a} \\
& = & \max_{\pi,C} E\left[ \int_0^T \chi f(C_s,J_s,s) ds + \varepsilon B (V_T^\pi) \right] - \lambda \left( E \left[ \int_0^T \chi \hat{C}_s ds + \hat{V}_T^\pi \right] - V_0 \right)  \nonumber \\
& = & \max_{\pi,C}  E\Bigg[  \chi \int_0^T  \left( f(C_s,J_s,s) - \lambda \hat{C}_s \right)  ds + \varepsilon B (V_T^\pi) - \lambda \left (\hat{V}_T^\pi - V_0 \right)  \Bigg] . \label{eq_Lagrange1c} 
\end{eqnarray}
Note that we move in \eqref{eq_Lagrange1a} the constraint that $\hat{G}^\pi$ has to be fair, under a single, overarching expectation. In some sense, if we are able to maximize the random variable inside the expectation in \eqref{eq_Lagrange1c}, then we have a candidate for the optimal payout, and thus, a clue for the optimal strategy.

To determine the optimal investment and consumption strategy, one is tempted to fix a time $s \in [0,T]$ and then determine the optimal consumption level at that point in time. With time-additive utility this approach finds the correct characterization of the consumption strategy, see \citet{Pen:08}. However, even with time-additive utility this approach ignores that setting consumption at a time $t$ affects the wealth to be invested going forward in time and, therefore, does impact future consumption choices. With Epstein-Zin preferences (recursive utility) these choices are even more intertwined, since future consumption recursively affects current utility levels in addition to consumption levels. To make such a calculation mathematically rigorous requires a suitably defined derivative of the entire consumption path over the time interval $[0,T]$.

In a series of papers Darrel Duffie and coauthors looked into this problem, see \citet{Duf&Eps:92} and \citet{Duf&Ski:94}, and used the mathematical concept of Gateaux-derivatives applied to consumption paths to show that these drive the SDF. Section 9.H in \citet{Duf:01} calculates the Gateaux-derivative explicitly for time-additive utility and shows that the popular characterization of an SDF holds in this case. With stochastic differential utility (Epstein-Zin preferences), \citet{Duf&Ski:94} characterize on pages 125-127 explicitly the SDF, see also Section 9.H together with Appendix F in \citet{Duf:01}. At any point in time their arguments allow us to derive from \eqref{eq_Lagrange1c} in our more general setting the following relations for the candidates of the optimal $V_T^*$ and $C_s^*$:
\begin{eqnarray}
\frac{dB }{dV} (V_T^*) & = & \frac{\lambda}{\varepsilon V_T^{GP}} ,
\end{eqnarray}
and, as long as $\chi=1$, for $0 \leq s < T$, and given $J_s$ one gets
\begin{eqnarray}
D_s \frac{\partial f}{\partial C} (C_s^*,J_s,s)  & = & \frac{\lambda}{V_s^{GP}} , \label{eq_ut_grad} 
\end{eqnarray}
where\footnote{With recursive preferences, the process $D$ captures the fact that marginal changes in consumption at any time affect (recursively) the entire consumption path and lifetime utility. With time-additive preferences, it can be shown that \eqref{eq_ut_grad} simplifies to the usual formula, where marginal utility is proportional to an SDF.}
\begin{eqnarray}
D_s & = & \exp\left( \int_0^s \frac{\partial f}{\partial J} (C_t^*,J_t,t) dt \right) , \label{eq_defn_proc_D}
\end{eqnarray}
see Example 3 on page 120 in \citet{Duf&Ski:94}.

For simplicity, we write $f^\prime(C,l,s)=\frac{\partial f}{\partial C} (C,l,s)$ and $B^\prime (V)=\frac{dB }{dV}(V)$.
Assumption \ref{ass2} implies that both functions $f'$ and $B'$ are invertible with respect to $C>0$ and $V>0$, respectively, and we denote by $f^{\prime,-1} (\cdot,l,s)$ for given $(l,s)$ and by $B^{\prime,-1}(\cdot)$ their respective inverse functions. This allows us to write the candidate for the optimal terminal wealth as
\begin{eqnarray}
V_T^* & = & B^{\prime,-1} \left( \frac{\lambda}{\varepsilon V_T^{GP}} \right) , \label{eq_opt_plan1}
\end{eqnarray}
and the candidate for the optimal consumption at time $s \in[0, T]$ as 
\begin{eqnarray}
C_s^* = \chi f^{\prime,-1} \left( \frac{\lambda}{D_s V_s^{GP}},J_s,s \right) . \label{eq_opt_plan2}
\end{eqnarray}

Note that the process $D$ in \eqref{eq_defn_proc_D} depends on the utility process $J$ and on the optimal consumption process $C^*$.
The key observation is here that the optimal consumption \eqref{eq_opt_plan2} and the terminal wealth characterization in (\ref{eq_opt_plan1}), as well as, the process $D$ depend only on the process $J$ and on the GP. 

At time $t=0$, the investor starts with initial wealth $V_0>0$. The Lagrange multiplier $\lambda$ must, therefore, solve equation \eqref{eq_value} at time $t=0$, using the optimal plan \eqref{eq_opt_plan1} and \eqref{eq_opt_plan2}. The Inada conditions in Assumption \ref{ass2} let us study the cases $\lambda \to 0$, and $\lambda \to \infty$. This shows us that the right-hand side in \eqref{eq_value} runs from $\infty$ to $0$, which ensures the existence of $\lambda$.

The well-known martingale technique for time-additive utility, see, e.g. Section 12.4.2 in \citet{Pen:08} or Section 4.4.3 in \citet{Cvi&Zap:04}, as well as its generalization to recursive preferences using the utility gradient technique, see Section 9.H in \citet{Duf:01}, all assume classical no-arbitrage assumptions and end up with the key observation that optimal consumption and terminal wealth can be represented in terms of the classical SDF. Our derivations here generalize this insight in a practically important direction: The SDF is the inverse of the tradeable GP, and our approach has the key advantage that we do not request the restrictive no-arbitrage assumptions of classical finance theory. Furthermore, in our approach the SDF is fully linked to tradeable securities, these are the GP and the baseline security. 

As we will show in the next section, these advantages are crucial from a practical point of view, as they have important implications for implementing the optimal strategy and ultimately for employing more realistic long term market models in managing risk than available under classical assumptions.

Let us summarize our findings in (\ref{eq_value})-(\ref{eq_opt_plan2}) using \eqref{setup_opt2} as follows: 

\begin{corollary}
Assuming that the optimization problem \eqref{eq_Lagrange1c} has a unique solution, then the candidates for the benchmarked optimal consumption-savings process $\hat{V}^*$ and for the indirect utility process $J$ are determined for $0 \leq t \leq T$ by the conditional expectations
\begin{equation}
\hat{V}_t^* = E \left[ \left. \chi \int_t^T f^{\prime,-1} \left( \frac{\lambda}{D_s V_s^{GP}},J_s,s \right) \frac{1}{V_s^{GP}} ds + B^{\prime,-1} \left( \frac{\lambda}{\varepsilon V_T^{GP}} \right)  \frac{1}{V_T^{GP}} \right| \mathcal{A}_t \right],  \label{eq_defn_opt_wealth} 
\end{equation}
and
\begin{equation}
J_t = E \left[ \left. \chi \int_t^T f \left( f^{\prime,-1} \left( \frac{\lambda}{D_s V_s^{GP}},J_s,s \right) ,J_s , s \right) ds + \varepsilon B \left( B^{\prime,-1} \left( \frac{\lambda}{\varepsilon V_T^{GP}} \right)  \right) \right| \mathcal{A}_t \right] , \label{eq_defn_opt_wealth2}
\end{equation}
respectively.
\end{corollary}
Note that the value on the right hand side of the benchmarked optimal consumption-savings process assumes neither the existence of an equivalent risk-neutral probability measure nor market completeness. The key assumption made is the existence of the GP, which needs trivially to be satisfied to avoid economically meaningful arbitrage.

\section{Trading the Optimal Consumption-Savings Process}\label{sec_attain}

The previous section found that the optimal consumption decision (at any time) and the terminal wealth depend ultimately \emph{only} on the process characterizing the GP in its given denomination, see equations \eqref{eq_opt_plan1} and \eqref{eq_opt_plan2}. Therefore, for further analysis, this section discusses modeling that process. 

Under the classical no-arbitrage paradigm, the vector of state variable processes, which models the entire market dynamics, determines, in general, also the optimal solution. Due to the enormous number of state variables that characterize the entire market dynamics, this is not a realistic way of implementing optimal strategies in practice and would leave our previous statements on a purely theoretical level. We have to face, in practice, the impossibility to model and estimate sufficiently accurate the dynamics of all components of the entire global market to implement useful optimal portfolios. This impossibility has been explained, e.g. in \citet{DeM&Gar&Upp:09} for the closely related task of sample based mean-variance portfolio optimization, and is also argued in \citet{Kan&Wan&Zho:16}, \citet{Kan&Zho:07}, and \citet{Okh&Scm:06}.

Instead of aiming for an extremely complex, purely theoretical model for the entire market with unresolvable modeling and parameter estimation challenges, we propose in this paper to exploit the above clarified central role of the GP for the characterization and construction of optimal portfolios. Therefore, it turns out to be extremely useful that proxies of the GP for a given investment universe can be constructed, as demonstrated in \citet{Pla&Ren:12}, and \citet{Pla&Ren:16}. This makes it then practically feasible to approximate well the targeted optimal portfolios and consumption processes. 

%In addition, we assume that the process $L$ in the preference specification \eqref{setup_opt} is given and fulfills the assumption:
%
%\begin{assumption}\label{ass_L}
%The process $L$ is Markovian and adapted to the filtration generated by the GP.
%\end{assumption}
%
%In the case where the process is endogenous, e.g. when it is an indirect utility function (as with recursive preferences) we will discuss later that this property if fulfilled such that the structural analysis in this section can applied as well. 

\subsection{Multi-Dimensional Markovian Models}

Multi-dimensional, continuous Markovian market dynamics appear to be the class of market models that have been implemented most successfully in the context of utility maximization and derivative pricing. For obtaining tractable optimal strategies we make, therefore, the following assumption:

\begin{assumption}\label{ass_GP}
The value $V_t^{GP}=V^{GP}(t,M_t^1,\ldots, M_t^n)$ of the discounted GP is a function of a multi-dimensional Markov process $M=\{ M_t=(M_t^1,\ldots, M_t^n)^\top, t \in [0,T]\}$. The vector process $M$ satisfies the SDE 
\begin{equation}
d M_t = \mu_{M} (t,M_t) dt + \sigma_{M} (t,M_t) d W_t, 
\end{equation}
where $\mu_M$ and $\sigma_M$ are suitable functions of time $t$ and the Markovian state vector $M_t$.
\end{assumption}

As discussed above, our analysis draws our attention to properties of the GP process. It is important to stress that assumption \ref{ass_GP} requires the drift vector and volatility matrix to be functions of time and of the vector process $M$, only. Although state variables may drive the drift and volatility of primary securities, these assumptions mean that no state variable drives the vector process $M$. In Section \ref{sec_emp} we will discuss a model that has this convenient property and matches empirically well the observed GP dynamics.

We assume from now on that we have a tradeable proxy of the GP, which we then identify with the GP for the purposes of this paper. For an example of a construction of a proxy for the GP of the global equity market we refer to \citet{Pla&Ren:12} and for the GP of developed equity markets to \citet{Pla&Ren:16}. Ultimately, we will show that the optimal consumption-savings process and investment strategy become tradeable in terms of suitable multiple funds and the baseline risk-free security, along the lines of well-established multiple-fund theorems, see e.g. \citet{Mer:71} or \citet{Pen:08}. Our analysis in Section \ref{sec_choice} postpones the issue of tradeability of optimal solutions but our discussion here now confirms the feasibility of our approach also in this respect.

The characterization of the bequest $B$ and the consumption $C$ in equations (\ref{eq_opt_plan1}) and (\ref{eq_opt_plan2}) tells us that these are driven by our vector Markov process. Furthermore, due to the characterization in equation \eqref{eq_defn_proc_D}, the process $D$ can be interpreted as a component of our vector Markov process. Finally, together with \eqref{eq_defn_opt_wealth2} and Assumption  \ref{ass_GP}, this implies that the recursive utility $J$ is interpretable as a component of our vector Markov process. Based on equations (\ref{eq_ut_grad}), (\ref{eq_defn_opt_wealth}) and (\ref{eq_defn_opt_wealth2}) we then conclude that the optimal wealth process $V^*$, the recursive utility process $J$, the consumption process $C^*$ and the process $D$ are all fully characterized by the current time $t$ together with the current values of the components of the Markov process $M$.

This allows us to introduce the optimal value (function) $V_t^*=V^*(t,M_t)$ and the life-time utility (function) $J_t=J(t,M_t)$ through equations (\ref{eq_defn_opt_wealth}) and (\ref{eq_defn_opt_wealth2}), as well as the function $D_t=D(t,M_t)$  through equation \eqref{eq_defn_proc_D}. This means, for $0 \leq t \leq T$ we have
\begin{eqnarray}
V^*(t,m) & = & V^{GP} (t,m) E \Bigg[ \chi \int_t^T \frac{C^* (s,M_s)}{V^{GP} (s,M_s)}  ds + \frac{V^* (T,M_T)}{V^{GP}(T,M_T)}  \Bigg| 
M_t =m 
\Bigg] , \label{eq_defn_Uhat} \\
J (t,m) & = & E \Bigg[ \chi \int_t^T f \left( C^* (s,M_s) , J (s,M_s),s \right)  ds + \varepsilon B (V_T^*) \Bigg| 
M_t =m 
\Bigg] ,  \label{eq_defn_Uhat2}
\end{eqnarray}
where
\begin{eqnarray}
V_T^* = V^* (T,M_T) & = & B^{\prime,-1} \left( \frac{\lambda}{\varepsilon V^{GP} (T,M_T)} \right) , \label{eq_BC0}
\end{eqnarray}
and, for $0 \leq s \leq T$ we get
\begin{eqnarray}
C^* (s,M_s) & = & \chi f^{\prime,-1} \left( \frac{\lambda}{ D (s,M_s) V^{GP} (s,M_s)} ,J (s,M_s),s \right) .  \label{eq_BC00}
\end{eqnarray}

Assuming sufficient differentiability of $V^*(\cdot, \cdot)$, an application of It\^o's lemma yields the SDE for $V_t^*=V^*(t,M_t) = V^*(t,M_t^1, \ldots, M_t^n)$ in the form
\begin{eqnarray}
dV_t^* & = & \left( \frac{\partial V^*}{\partial t} + \frac{1}{2} \sum_{i,j=1}^n \sigma_{Mi} \sigma_{Mj}  \frac{\partial^2 V^*}{\partial m_i \partial m_j} \right) dt + \sum_{i=1}^n \frac{\partial V^*}{\partial m_i} d M_t^i . \label{eq_opt_wealth1_gen} 
%& = & V_t^* (a_t^V dt + b_t^{V} d W_t) , \label{eq_opt_wealth2_gen} 
\end{eqnarray}
The terms in front of $dM_t^i$ in equation \eqref{eq_opt_wealth1_gen} yield by standard hedging arguments the following insight:

\begin{corollary}[Multiple Fund Separation]\label{tfsep_thm_gen}
Assume that all components $M^i,i=1,\ldots,n$ of $M$ are constructed in such a way that they can be traded, giving rise to $n$ risky non-redundant funds plus the baseline security. Then the investor can implement her optimal investment strategy (and fund her consumption-savings profile) by holding at time $t$
\begin{equation}
\frac{\partial V^*}{\partial m_i} \label{eq_thm_gen_fund}
\end{equation}
units of the $i$-th fund $M_t^i, i=1,\ldots,n$, and invest the remainder of her wealth in the given baseline security, the locally risk-free security.
\end{corollary}

Our proof of this result is mathematically similar to those of prior multiple fund separation theorems in the consumption-savings literature, see e.g. \citet{Pen:08}. This literature suggests that additional state-variables give rise to additional funds that play a necessary role in optimal investing. Our results, however, clarify this by placing the emphasis fully on the GP denominated in the baseline security and, thus, on the components of $M$ that determine its dynamics. \emph{Only} these components are needed for constructing an optimal investment portfolio and not the many other quantities that characterize the entire financial market. Furthermore, we allow a significantly richer modeling world for capturing more realistically the long-term dynamics of the GP than classical finance theory permits. We request only the existence of the GP and make no longer the additional assumption on the existence of an equivalent risk-neutral probability measure. Since we have discounted our securities by the locally risk-free baseline security, we have to model also its dynamics in our Markovian system, when targeting payoffs that refer to currency units or inflation adjusted payouts, which is a standard task.

It is important to note that in our approach one does not have to care about market incompleteness, i.e. about any additional factors that may be needed to complete the market. All that is needed are the $n$ funds that arise from the $n$ components of $M_t$. These characterize $V_t^*$ and no other funds are necessary. This reduces the task of portfolio optimization to its core and clarifies the natural inputs that determine the optimal strategy with the GP as central building block.

In addition, our clarification simplifies considerably practical portfolio construction, in particular, when a proxy of the GP is employed. The number of factors needed to model the GP is significantly smaller than the number of state variables characterizing an entire global market model. 
This becomes most evident, when the uncertainties driving the GP can be captured in a single Brownian motion. In the case of the equity market, the GP is then driven by one source of uncertainty, called the non-diversifiable (systematic) uncertainty. Despite its simplicity, this appears to be a rather realistic case, as forthcoming research will reveal. We will elaborate on a stylized version of this case in Section \ref{sec_emp} and then show how this insight simplifies practically relevant applications.

The reader accustomed with optimal investing may be suprised to see that, according to Corollary \ref{tfsep_thm_gen}, so-called intertemporal hedge demands do not arise but that only positioning in the terms driving the GP matters. Two important observations must be made: First, we note that it may well be that intertemporal hedge demand show up in other representations of demand that look at positioning in the primary securities instead of positioning in (the driving forces of) the GP. Most important, we note that Assumption \ref{ass_GP} does not permit the Markov process $M$ to be driven by state variables that are not spanned by the primary securities. When the vector Markovian process $M$ would be driven by unspanned state variables, on might well obtain some intertemporal hedge demand. While one can easily imagine situations where intertemporal hedge demands apear, the scope of this paper is to focus on practical feasibility and point out practically relevant situations where the analysis can be simplified. As such we are not interested in analyzing situations itself that lead to more complex dynamic asset allocation.
                
\subsection{Scalar Markovian Growth Optimal Portfolio}\label{subsec_scalar_GP}

A particularly important case results when the GP forms a scalar Markov process, that is when we have $n=1$ and can aggregate its driving uncertainty in a scalar Brownian motion $W$. Then, we can replace $M_t$ by the GP value $V_t^{GP}$ at time $t$ and write $V_t^*$ as a function of the current GP value, i.e. $V^* = V^* (t,V_t^{GP})$. 

Assuming sufficient differentiability of the function $V^*$, an application of It\^o's lemma yields the SDE for $V_t^*=V^*(t,V_t^{GP})$. We then obtain by equation \eqref{eq_dyn_GP} that 
\begin{eqnarray}
dV_t^* & = & \left( \frac{\partial V^*}{\partial t} + \frac{1}{2} \theta_t^2 (V_t^{GP})^2 \frac{\partial^2 V^*}{\partial v^2} \right) dt + \frac{\partial V^*}{\partial v} d V_t^{GP} \label{eq_opt_wealth1} \\
& = & V_t^* (a_t^V dt + b_t^{V} d W_t) , \label{eq_opt_wealth2} 
\end{eqnarray}
where
\begin{eqnarray}
a_t^V V_t^* & = & \frac{\partial V^*}{\partial t} + \frac{1}{2} \theta_t^2 \left(V_t^{GP} \right)^2 \frac{\partial^2 V^*}{\partial v^2} + \theta_t^2 V_t^{GP} \frac{\partial V^*}{\partial v}, \label{eq_opt_diff_terms0} \\
\text{ and } \quad
b_t^V V_t^* & = & \theta_t V_t^{GP} \frac{\partial V^*}{\partial v}   .  \label{eq_opt_diff_terms}
\end{eqnarray}
The term in front of $dV_t^{GP}$ in equation \eqref{eq_opt_wealth1} reveals the following result:

\begin{theorem}[Two-Fund Separation]\label{tfsep_thm}
When the discounted GP forms a scalar Markov process, the investor can implement her optimal investment strategy (and fund her optimal consumption-savings profile) by holding at all times 
\begin{equation}
\omega_t= \frac{\partial V^*}{\partial v} (t,V_t^{GP}) \label{eq_thm_2fund}
\end{equation}
units of the (risky) GP, and invest the remainder of her wealth in the given (riskless) baseline security. 
\end{theorem}

It is important to note that in this scalar Markovian case the optimal portfolio can be fully characterized through investment in the GP and the baseline security. This may come as a surprise, as the reader may be accustomed to wealth processes depending on various (untraded or additional) state variables. Here, however, there are no untraded or additional state variables that play any role in the optimal solution, as long as the baseline security is traded, which we assume here. Keeping in mind that our modeling is more general than modeling under classical no-arbitrage assumptions, Theorem \ref{tfsep_thm} generalizes various earlier results in the literature. For example, \citet{Pen:08} reports in his equation (12.70) the wealth weights with unspanned state variables. His equation coincides formally with ours, when all state variables are traded. Yet, our additional observations here are that the scalar Markovianity of the GP simplifies the strategy significantly to an investment into two funds. 
Note that a similar, slightly more general two-fund separation arises when only one Brownian motion drives a multi-component Markovian SDE that determines the GP as that of one of its components. In this case the equations \eqref{eq_opt_wealth1}-\eqref{eq_opt_diff_terms} become slightly more general when applying the It\^o formula to $V^*(t,M_t)$.

\section{The Optimal Portfolio Value}

The previous section expresses the investment strategy through first-order derivatives of the wealth function, i.e. as a function of the optimal portfolio value. This provides the important insight that it is sufficient to focus primarily on trading the GP and the baseline security. In practical applications this results in a crucial simplification when a proxy of the GP is employed. For potential further theoretical insights but mostly for practical implementations it remains to characterize further the process $V^*$. To simplify our exposition, we focus throughout this section on the case of a \emph{scalar} Markovian GP, discussed in Subsection \ref{subsec_scalar_GP}. The handling of more general Markovian multi-component models for characterizing the function $V^* (t,M_t)$ is then straightforward.

The first subsection explains how the general case should be addressed. The following two subsections explain how common preference specifications are covered within our preference framework of Subsection \ref{sec_PS_subsec_hedge}. In particular, they characterize the investment strategies that result for a scalar Markovian GP dynamics.

\subsection{The General Case}

The benchmarked consumption adjusted optimal wealth process $\hat{G}_t^* = \hat{V}_t^*+ \chi \int_0^t \hat{C}_s^* ds$, defined through equation \eqref{eq_Ftilde_help4}, fulfills the SDE
\begin{eqnarray*}
d \hat{G}_t & = & d \left(V_t^* \frac{1}{V_t^{GP}} \right) + \chi \hat{C}_t^* dt = V_t^* d \left(\frac{1}{V_t^{GP}}\right) + \frac{1}{V_t^{GP}} dV_t^* + d<V^*,\frac{1}{V^{GP}}>_t + \chi \hat{C}_t^* dt \\
& = & \frac{1}{V_t^{GP}} \left(   a_t^V V_t^* dt + \chi  C_t^*  - \theta_t  b_t^V V_t^* \right) dt + \left( \frac{V_t^*}{V_t^{GP}} b_t^V - \frac{V_t^*}{V_t^{GP}} \theta_t \right) d  W_t .
\end{eqnarray*}
This process has to satisfy equation \eqref{eq_Ftilde_help8}. As mentioned earlier, the process $\hat{G}$ must be a martingale. Consequently, we must have 
\begin{equation}
V_t^* \left( a_t^V - \theta_t b_t^V \right) + \chi C_t^* = 0. \label{eq_PDE0}
\end{equation}

Together with equations (\ref{eq_opt_diff_terms0}) and (\ref{eq_opt_diff_terms}) this allows us to formulate a partial differential equation (PDE) that characterizes $V^*(t,v)$ as follows:
\begin{equation}
\frac{\partial V^*}{\partial t} +  \frac{1}{2}  \theta_t^2 v^2 \frac{\partial^2 V^*}{\partial v^2} + \chi C^*(t,v) = 0. \label{eq_PDE1}
\end{equation}

Furthermore, we have for $V_t^*$ by \eqref{eq_dyn_GP} and \eqref{eq_thm_2fund} the SDE
\begin{equation}
dV_t^* = \left( \theta_t^2 V_t^{GP} \frac{\partial V^*}{\partial v} - \chi C_t^* \right) dt + \theta_t V_t^{GP} \frac{\partial V^*}{\partial v} dW_t = \omega_t d V_t^{GP} - \chi C_t^* dt , \label{eq_PDE2n}
\end{equation}
see Theorem \ref{tfsep_thm}. Since $C^*$ depends on $J^*$, see equation \eqref{eq_BC00}, in general, the PDE \eqref{eq_PDE1} needs to be solved jointly with the Hamilton-Jacobi-Bellman (HJB) PDE for $J$. Equation (3) of Section 9.A in \citet{Duf:01} reports the HJB equation for a controlled process given in his equation (1). In our setup, equation  \eqref{eq_PDE2n} specifies the controlled process and we find that the HJB equation reads
\begin{equation}
\max_{C^*(t,v)} \left\{ \frac{\partial J^*}{\partial t} + \left( \theta_t^2 \frac{\partial V^*}{\partial v} v- \chi C_t^* \right) \frac{\partial J^*}{\partial V^*}+ \frac{1}{2} \theta_t^2 v^2 \left( \frac{\partial V^*}{\partial v} \right)^2 \frac{\partial^2 J^*}{\partial V^{*2}} + \chi f \left( C^*, J^* ,t \right) \right\} = 0 . \label{eq_PDE2prev}
\end{equation}
When $\chi=1$ (consumption-savings problem), this means that (formally) the first-order condition yields
\begin{equation}
- \frac{\partial J^*}{\partial V^*} + f'(C^*,J^*,t)=0,
\label{eq_PDE2new}
\end{equation}
where
\begin{equation}
C^* (t,v)= f^{\prime,-1} \left(\frac{\partial J^*}{\partial V^*}, J^* ,t \right) .
\end{equation}
Using this characterization together with equations \eqref{eq_PDE1} and \eqref{eq_PDE2prev} we have to solve the system of PDEs
\begin{eqnarray}
\frac{\partial V^*}{\partial t} +  \frac{1}{2}  \theta_t^2 v^2 \frac{\partial^2 V^*}{\partial v^2} + \chi f^{\prime,-1} \left(\frac{\partial J^*}{\partial V^*}, J^* ,t \right)  & = & 0 , \\
 \frac{\partial J^*}{\partial t} + \frac{1}{2}  \theta_t^2 v^2 \left( \frac{\partial V^*}{\partial v} \right)^2 \frac{\partial^2 J^*}{\partial V^{*2}} + \chi f \left( f^{\prime,-1} \left(\frac{\partial J^*}{\partial V^*}, J^* ,t \right) , J^*, t \right) && \nonumber \\
 + \left( \theta_t^2 \frac{\partial V^*}{\partial v} v - \chi f^{\prime,-1} \left(\frac{\partial J^*}{\partial V^*}, J^* ,t \right) \right) \frac{\partial J^*}{\partial V^*} & = & 0 . \label{eq_PDE2}
\end{eqnarray}
Based on equation \eqref{eq_BC0} and \eqref{eq_defn_Uhat2} it remains to satisfy also the terminal boundary conditions
\begin{equation}
V^* (T,v)= B^{\prime,-1} \left( \frac{\lambda}{\varepsilon v} \right) 
\text{ and } J^* (T,v) = \varepsilon B \left( B^{\prime, -1} \left( \frac{\lambda}{\varepsilon v} \right) \right) , \label{eq_BC1}
\end{equation}
respectively. Finally, to assure the martingale property for $\hat{G}$ we impose the following spatial boundary conditions: For $v \to \infty$ we require $V^*(t,v) \to \infty$ and for $v \to 0$ we require $V^*(t,v) \to 0$.

Note that the solution depends on the Lagrange multiplier $\lambda$, i.e. the solution to the PDE \eqref{eq_PDE1} is characterized by $\lambda$. This parameter should be set to fulfill the initial budget constraint, see our previous discussion in Section \ref{sec_choice}. Since the characterization of the optimal portfolio value depends strongly on the choice of $f$, the next subsections discuss particular cases. 

\subsection{(Time-additive) Preferences over Consumption}\label{subsec_pref1}

In (time-additive) consumption savings problems ($\chi=1$) with a given utility function $u$ and a rate of time-preference $\delta>0$ we set\footnote{Alternatively, but more in line with the literature, one may consider the aggregator as $f(c,l,t)=u(c) - \delta l$ to also find that $D_s \frac{\partial f}{\partial c} = e^{-\delta s} u'(c)$. The literature on stochastic differential utility shows that this functional form of the aggregator leads to a $J$ process that is equivalent in terms of preference ranking to the common specification $J_t=\max_{\pi,C} E[\int_t^T \exp(- \delta (T-s)) u(C_s) ds | \mathcal{A}_t]$, see \citet{Duf&Eps:92}.} $f(c,l,t) = e^{-\delta t} u(c)$. We then obtain $D=1$ by equation \eqref{eq_defn_proc_D} and $D_s \frac{\partial f}{\partial c} = e^{-\delta s} u'(c)$ for the term on the left-hand side in equation \eqref{eq_ut_grad}. 

The literature often illustrates time-separable consumption-savings problems with CRRA preferences. Using a (constant) rate of time-preference  $\delta>0$ and a risk-aversion coefficient $0 < \gamma$, one looks typically at $B: x \mapsto e^{-\delta T} x^{1-\gamma}/(1-\gamma)$ and $f:(c,t) \mapsto e^{-\delta t} c^{1-\gamma}/(1-\gamma)$ for $\gamma \neq 1$. For $\gamma=1$ the functions are analogous, replacing the power functions by the natural logarithm and leading to $B: x \mapsto e^{-\delta T} \ln (x)$ and $f:(c,t) \mapsto e^{-\delta t} \ln (c)$. Our preference specification covers these cases of CRRA preferences and several more general cases that fulfill Assumption \ref{ass2}. 

For further illustration throughout this subsection we discuss exclusively the case of time-additive utility, where the investor has CRRA preferences, as described above. 
We then calculate $f^{\prime,-1}(c,l,s)=(e^{-\delta s} c)^{-1/\gamma}, B^{\prime,-1}(x)=(e^{\delta T} x)^{-1/\gamma}$, which gives
\begin{eqnarray}
C_s^* & = & (e^{\delta s} \lambda)^{-1/\gamma} (V_s^{GP})^{1/\gamma} , \text{ for } 0 \leq s < T , \text{ and } 
V_T^* = \varepsilon^{1/\gamma} (e^{\delta T} \lambda)^{-1/\gamma}   \left( V_T^{GP} \right)^{1/\gamma} .  \label{eq_CRRA_cons}
\end{eqnarray}
This yields
\begin{eqnarray*}
\frac{V^* (t,v)}{v} & = & \frac{1}{\lambda^{1/\gamma}} \int_t^T  e^{- \frac{\delta}{\gamma} s} E \left[ \left.  (V_s^{GP}) ^{(1/\gamma)-1} \right| V_t^{GP} =v \right] ds \\
& & +  \frac{\varepsilon^{1/\gamma} e^{- \frac{\delta}{\gamma} T}}{\lambda^{1/\gamma}} E \left[ \left. (V_T^{GP})^{(1/\gamma)-1} \right| V_t^{GP} =v \right] .
\end{eqnarray*}
We could characterize the consumption-wealth ratio $C_t^* / V^* (t,v)$, but refrain from doing so, since our focus is on investment strategies. 
The initial budget equation $V_0=V_0^*=V^* (0,V_0^{GP})$ with $V_0^{GP}=1$ sets $\lambda$, 
which in turn gives the benchmarked portfolio value as
\begin{eqnarray}
\frac{V^* (t,v)}{v} & = & V_0 \frac{E \left[ \left. \chi \int_t^T e^{-(\delta/\gamma) s} (V_s^{GP})^{(1/\gamma)-1} ds + \varepsilon e^{-(\delta/\gamma )T} (V_T^{GP})^{(1/\gamma)-1} \right |V_t^{GP} =v \right]}{E \left[ \chi  \int_0^T e^{-(\delta/\gamma) s} (V_s^{GP})^{(1/\gamma)-1} ds + \varepsilon e^{-(\delta/\gamma) T} (V_T^{GP})^{(1/\gamma)-1}\right]}  \label{eq_CRRA_Vstar} 
%& = & V_0 v^{(1/\gamma)-1} \frac{E \left[ \left. \chi \int_t^T e^{-(\delta/\gamma) s} \left( \frac{V_s^{GP}}{v} \right)^{(1/\gamma)-1} ds + \varepsilon e^{-(\delta/\gamma )T}\left ( \frac{V_T^{GP}}{v} \right)^{(1/\gamma)-1} \right |V_t^{GP} =v \right]}{E \left[ \chi  \int_0^T e^{-(\delta/\gamma) s} (V_s^{GP})^{(1/\gamma)-1} ds + \varepsilon e^{-(\delta/\gamma) T} (V_T^{GP})^{(1/\gamma)-1}\right]} \ . \nonumber
\end{eqnarray}
Taking the derivative gives according to \eqref{eq_thm_2fund} 
\begin{equation*}
\omega_t= \frac{\partial V^* (t,v)}{\partial v} .  \label{eq_CRRA_holding}
\end{equation*}
Recall from Theorem \ref{tfsep_thm} that this gives the number of units of the GP to be held and that $v$ refers to the current value of the GP. 

Specifying the stochastic dynamics of the GP, we could calculate  for $t \leq s \leq T$ the expectations $E[(V_s^{GP})^{(1/\gamma)-1}|V_t^{GP} =v]$ and determine the optimal value function $V^*$. This would then characterize the optimal investment strategy through Theorem \ref{tfsep_thm}. We will discuss this further in the next section, where we consider several dynamics for the GP.

A particularly convenient case is when the optimal value function does not depend on expectations $E[(V_s^{GP})^{(1/\gamma)-1}|V_t^{GP} =v]$: This corresponds to the case of a log-investor ($\gamma=1$). Then, the above equations simplify to
\begin{eqnarray}
V_t^* & = & V_0 \frac{\varepsilon \delta + e^{-\delta (t-T)} -1 }{\varepsilon \delta + e^{\delta T} -1 } V_t^{GP}  . \label{eq_ln_solve_V} 
\end{eqnarray}
Additional calculations would recover equations, shown e.g. in \cite{Pen:08}.

Equation \eqref{eq_ln_solve_V} shows that the optimal value process grows similarly to the GP, where a time-dependent function determines the effect of consumption. This suggests that the investor does not hold all her wealth in the GP. We then calculate $\frac{\partial V^*}{\partial v} = \frac{V_t^*}{v}$
and find based on equation \eqref{eq_thm_2fund} for a scalar Markovian GP that the investor holds at time $t \in [0,T]$:
\begin{equation*}
V_0 \frac{\varepsilon \delta + e^{-\delta (t-T)} -1 }{\varepsilon \delta + e^{\delta T} -1 } 
\end{equation*}
units of the GP and the remainder in the locally riskless asset. At the terminal time $T$ the bequest amounts to $V_T^* = V_T^{GP} V_0 \frac{\varepsilon \delta }{\varepsilon \delta + e^{\delta T} -1 }$. 

Note that as we increase $T$, the log-investor holds more and more of her wealth in the GP; taking the limit $T \to \infty$ we find that she holds all her wealth in the GP, which (formally) matches the earlier mentioned insight on the Kelly portfolio.

We emphasize that in the case of a log-investor the optimal strategy is independent of the dynamics of the GP, which is an extremely important observation because the log-utility maximizing portfolio is the one that in the long run generates almost surely the highest value, see Theorem 10.5.1 in \citet{Pla&Hea:10}. Thus, an investor who prefers more for less and has an extremely long time horizon is naturally behaving as a log-investor. This type of investor is typically also deeply concerned about the sustainability of our consumption and economic activity, which has become more and more the focus of attention.

\subsection{Preferences over Terminal Wealth}\label{subsec_pref0}

In asset allocation problems ($\chi=0$) it is customary to introduce a rate of time-preference $\delta>0$, a risk-aversion coefficient $\gamma >0$, and set $f=0$, as well as $B: x \mapsto \exp( -\delta T) x^{1-\gamma}/(1-\gamma)$ for $\gamma \neq 1$ and $B:x \mapsto \exp( - \delta T) \ln (x) $ for $\gamma=1$. Our preference specification covers these and several more general cases of preferences that fulfill Assumption \ref{ass2}. 

Preferences over terminal wealth are a special case of our analysis in the previous subsection. For completeness and comparison with the literature, we report the results from equations \eqref{eq_CRRA_cons} and \eqref{eq_CRRA_Vstar}:
\begin{eqnarray}
V_T^* & = & \varepsilon^{1/\gamma} (e^{\delta T} \lambda)^{-1/\gamma}   \left( V_T^{GP} \right)^{1/\gamma} \label{eq_TW_holding0} \\
\frac{V^* (t,v)}{v} & = & V_0 \frac{E \left[\left.  \left( V_T^{GP}\right)^{(1/\gamma)-1} \right|V_t^{GP}=v \right]}{E[ (V_T^{GP} )^{(1/\gamma)-1}]} \ . \label{eq_TW_value} 
\end{eqnarray}
As before, we recall from Theorem \ref{tfsep_thm} that $\frac{\partial V^*}{\partial v}$ gives the number of units of the GP to be held and that $v$ refers to the current value of the GP. The results in equations \eqref{eq_TW_holding0}-\eqref{eq_TW_value} can also be calculated directly, setting $f=0$ such that $D=1$ according to equation \eqref{eq_defn_proc_D}, and calculating $B^{\prime,-1}(x)=(e^{\delta T} x)^{-1/\gamma}$.

Equations \eqref{eq_TW_holding0}-\eqref{eq_TW_value} simplify considerably when we assume logarithmic preferences, that is $\gamma=1$. In that case we find
\begin{equation}
\frac{V^* (t,v)}{v} = V_0 ,
\end{equation}
i.e. the investor holds exclusively the GP, as mentioned already in Subsection \ref{rw_price}. This is the case where for $T \to \infty$ one obtains almost surely the highest portfolio value. More precisely, on obtains the largest long-term growth rate, that is
$$
\lim_{T \to \infty} \ln \left( \frac{V_T^{GP}}{V_0} \right) \geq \lim_{T \to \infty} \ln \left( \frac{V^*( T, V_T^{GP})}{V_0} \right)
$$
$P$-almost surely, by the GP: This key property of the GP makes it so special among all other optimal portfolios and explains intuitively its central role in portfolio optimization.

\section{An Empirical Evaluation}\label{sec_emp}

The previous sections introduced our approach based on several assumptions, in particular continuous trading, tradeability of the SDF and the Markov property of its process. This allowed us to come up with convenient consumption-savings and asset allocation strategies that should lead to superior wealth levels. Ultimately, however, it is an empirical question whether our approach does lead to superior wealth levels when implemented in practice. This section aims to provide support for this.

We study an investor who wants to invest in the US stock market over the long run. The S\&P500 total return index is employed for the time period from 1925 until today; this reconstructed index is provided by \citet{Shi:15}. Throughout this section we focus on subperiods of the time from 1925 to 2017 and identify the GP with the S\&P500.  

As noted at various stages throughout this paper, preferences with constant relative risk aversion (CRRA) are a common assumption in financial modeling. Therefore, we also adopt these for empirical evaluation. For simplicity, we evaluate terminal wealth, i.e. we set
$f=0, \chi=0, \ \varepsilon=1$ and $B: x \mapsto \exp( -\delta T) x^{1-\gamma}/(1-\gamma)$ for $\gamma \neq 1$, respectively $B:x \mapsto \exp( - \delta T) \ln (x) $ for $\gamma=1$. As usual, $\gamma$ refers to the (relative) risk aversion coefficient. We are then in the asset allocation setup of Subsection \ref{subsec_pref0} and note that the investment strategy is well-specified through the number 
$$
\omega_t= \frac{\partial V^* (t,v)}{\partial v} 
$$
of units to hold in the S\&P500, and the remainder in the risk-free security. 
%Note that the popular advice of a 60/40 investment strategy then simply corresponds to the optimal strategy of an investor with relative risk aversion coefficient $\gamma=1/0.6 \approx 1.67$.

For implementations it remains to determine the value $V^*$ throughout time and states, which requires us to calculate the (conditional) expectations in equation \eqref{eq_TW_value}. Throughout we present results for a relative risk aversion coefficient $\gamma=3$, since this is a popular choice in finance. 

The process dynamics of the S\&P500 has been extensively researched and many process specifications in continuous-time have been studied. Throughout this paper we are particularly interested in Markovian processes for the GP driven by a single Brownian motion. An illustrative and rather realistic case arises when the GP of the stock market is assumed to follow the dynamics of the minimal market model (MMM) of \citet{Pla:01}, see also Chapter 13 in \citet{Pla&Hea:10}. When using the S\&P500 total return index as a proxy for the GP for the period 1871 to 2017, we obtain $\alpha_0= 0.1828$ and $\eta = 0.0520$. This allows us to calculate the above mentioned expectations $E \left[\left.  \left( V_T^{GP} \right)^{(1/\gamma)-1} \right|V_t^{GP}=v \right]$ for any time $t$. For completeness, details on the MMM and calculations are provided in Appendix \ref{app_MMM}.

\begin{figure}[ht!]
\begin{center}
\includegraphics[width=6in]{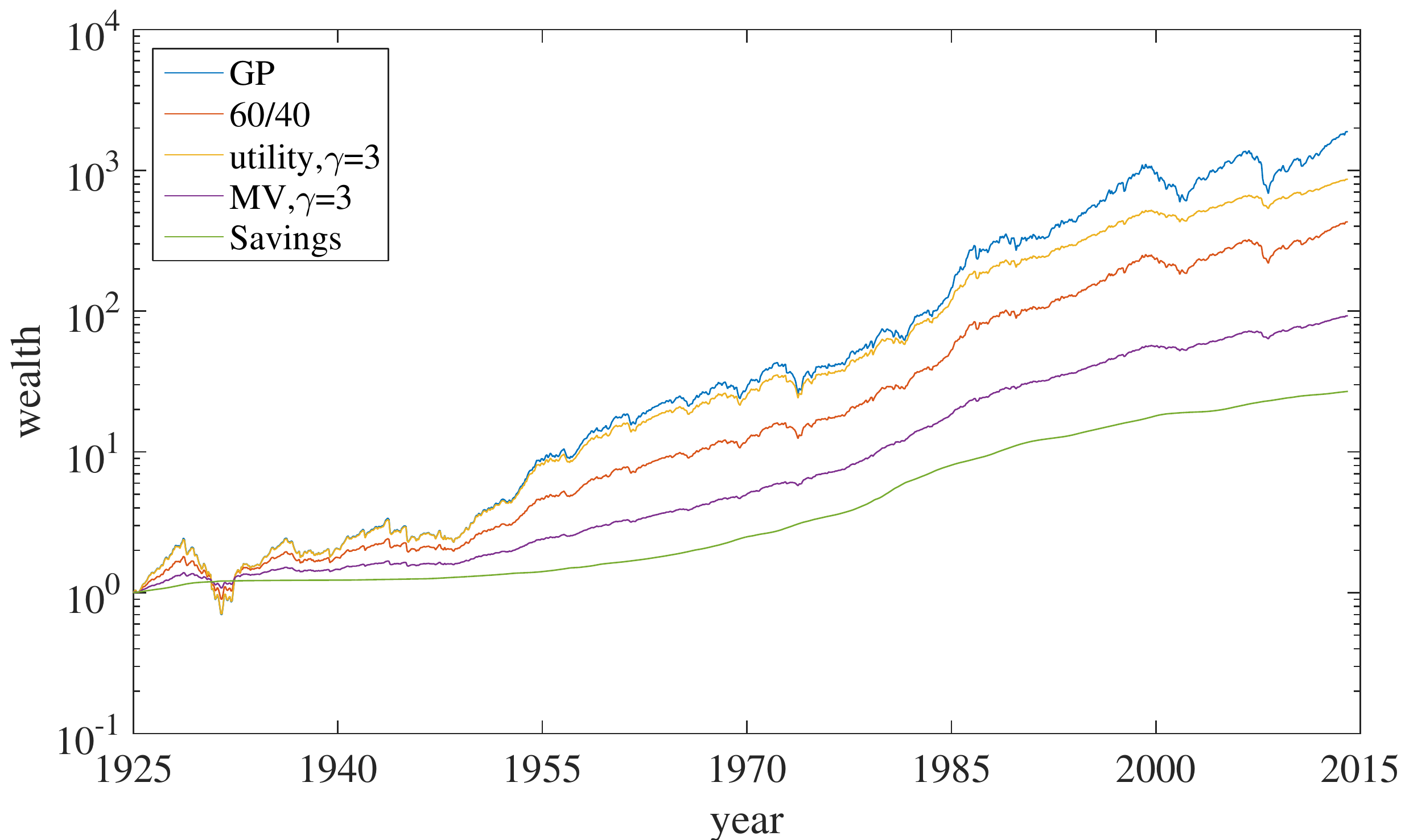}
\end{center}
%}
\caption{Wealth dynamics for different investment strategies from 12/1925 to 1/2015.}\label{sec_emp_fig1}
\end{figure}

We stress that we adopt the MMM process dynamics \emph{only} to allow us the calculation of the (conditional) expectations in equation \eqref{eq_TW_value} and the associated portfolio holdings in the S\&P500. Any other Markovian process characterization for the GP would also allow us to proceed analogously. Clearly, the better we capture the GP and the stochastic dynamics of the GP, the better our strategy will perform in implementation, however, the task of finding the appropriate proxy for the GP and a respective model is beyond the scope of this paper.

Figure \ref{sec_emp_fig1} presents the monthly evolution of wealth for our investor from 12/1925 to 1/2015, when she starts with initial wealth \$1 (yellow line). Her wealth would have grown considerably and increased to approximately \$865 under our optimal strategy, such that we plot wealth on a log scale for easier analysis. To put this into perspective, let us compare this with four alternative strategies. The first is the growth-optimal strategy for the long run, which invests all wealth in the S\&P500 and is the optimal strategy of a log-investor, see our earlier discussions; we plot this in Figure \ref{sec_emp_fig1} as a blue line. The growth-optimal portfolio leads to the highest growth in the long run such that it makes sense to invest largely in it in the initial years of a long-term investment strategy; as expected, our optimal portfolio closely tracks the GP initially, up to about 1955. The GP also comes with increased risk and so, as we come closer to the end date (2015), our optimal strategy departs from the GP, keeping a larger fraction of wealth in the risk-free security. The increased risk can be inferred from greater ``smoothness'' of our optimal portfolio value path across this almost 90 year time period, in particular since approximately 1985.

\begin{table}[ht!]
%\small
\begin{center}
\begin{tabular}{lrrrrrrrrr}
\toprule
& \multicolumn{9}{c}{end year} \\
\cmidrule(lr){2-10}
start \\
year & 1935 & 1945 & 1955 & 1965 & 1975 & 1985 & 1995 & 2005 & 2015 \\
\cmidrule(lr){1-1}
\cmidrule(lr){2-2}
\cmidrule(lr){3-3}
\cmidrule(lr){4-4}
\cmidrule(lr){5-5}
\cmidrule(lr){6-6}
\cmidrule(lr){7-7}
\cmidrule(lr){8-8}
\cmidrule(lr){9-9}
\cmidrule(lr){10-10}
\\
\multicolumn{10}{c}{Panel (a): Optimal Utility Strategy} \\
\cmidrule(lr){1-10}
1925 & 1.91 & 2.91 & 8.31 & 20.69 & 32.97 & 120.20 & 333.60 & 572.63 & 865.14 \\
1935 & &  1.52 & 4.35 & 10.82 & 17.24 & 62.86 & 174.47 & 299.49 & 452.47 \\
1945 & & &  2.86 & 7.11 & 11.33 & 41.31 & 114.66 & 196.82 & 297.36 \\
1955 & & & &  2.49 & 3.97 & 14.46 & 40.14 & 68.90 & 104.09 \\
1965 & & &  &  & 1.59 & 5.81 & 16.13 & 27.68 & 41.82 \\
1975 & & & &   &  & 3.65 & 10.12 & 17.37 & 26.24 \\
1985 & & & & & & & 2.78 & 4.76 & 7.20 \\
1995 & & & & & &   &  & 1.72 & 2.59 \\
2005 & & & & & & &  &  &  1.51 \\
\\
\multicolumn{10}{c}{Panel (b): 60/40 Strategy} \\
\cmidrule(lr){1-10}
1925  & 1.69 & 2.25 & 4.69 & 9.77 & 15.99 & 52.99 & 147.50 & 267.98 & 430.67 \\
1935 &  & 1.33 & 2.77 & 5.79 & 9.47 & 31.37 & 87.32 & 158.65 & 254.97 \\
1945 &  & & 2.09 & 4.35 & 7.12 & 23.58 & 65.64 & 119.26 & 191.67 \\
1955 &  & & & 2.09 & 3.41 & 11.31 & 31.47 & 57.17 & 91.89 \\
1965 &  & & & & 1.64 & 5.42 & 15.09 & 27.42 & 44.06 \\
1975 &  & & & &  & 3.31 & 9.22 & 16.76 & 26.93 \\
1985 &  & & & & & & 2.78 & 5.06 & 8.13 \\
1995 &  & & & & &  & & 1.82 & 2.92 \\
2005 &  & & & & & &  & & 1.61 \\
\\
\multicolumn{10}{c}{Panel (c): Mean-variance Strategy} \\
\cmidrule(lr){1-10}
1925 & 1.42 & 1.64 & 2.41 & 3.90 & 6.69 & 18.60 & 39.82 & 63.55 & 92.62 \\
1935 & & 1.15 & 1.69 & 2.74 & 4.70 & 13.06 & 27.95 & 44.62 & 65.02 \\
1945 & & & 1.47 & 2.38 & 4.09 & 11.36 & 24.33 & 38.83 & 56.58 \\
1955 & & & & 1.62 & 2.78 & 7.73 & 16.55 & 26.42 & 38.50 \\
1965 & & & & & 1.72 & 4.77 & 10.22 & 16.31 & 23.77 \\
1975 & & & & & & 2.78 & 5.95 & 9.49 & 13.83 \\
1985 & & & & & & & 2.14 & 3.42 & 4.98 \\
1995 & & & & & & & & 1.60 & 2.33 \\
2005 & & & & & & & & & 1.46 \\
\bottomrule
\end{tabular}
\end{center}
\caption{Total return varying start and end date.\label{table_prob}\label{table_prob2}}
\end{table}  

A second comparison is with the popular 60/40 strategy for the long run, which invests at all times 60\% (40\%) of her current wealth in the S\&P500 (the risk-free security). The red line in Figure \ref{sec_emp_fig1} plots the wealth of our investor following a 60/40 strategy with monthly rebalancing. The 60/40 strategy does not lead to a large decrease in value during the great depression but the GP and more important our strategy both pick up quickly and outperform the 60/40 strategy from thereon. This confirms that our strategy is intended for the long run.

A third comparison is with a monthly rebalancing investment strategy of a investor with mean-variance preferences. It is well-known that her portfolio weight in the risky asset (here the S\&P 500) is $\mu_t/(\sigma_t^2 A)$, where $\mu_t,\sigma_t$ denotes the conditional expectation and standard deviation of excess returns over the time period under consideration, as well as, $A$ denotes her risk aversion parameter. A straightforward Taylor approximation shows that in such a first approximation of the CRRA preferences with risk-aversion parameter $\gamma=3$ the risk-aversion of our mean-variance investor should be set to $A=2\gamma=6$. The unconditional parameters $\mu=6\%, \sigma=20\%$ are common choices in finance to model the S\&P500 annual return. Under usual time rescaling, the ratio $\mu/\sigma_t^2$ is independent of the time-horizon; given the well-known difficulties to implement the mean-variance approach conditionally we adopt these values and implement the mean-variance strategy unconditionally. We plot in Figure \ref{sec_emp_fig1} the wealth of our investor following the associated mean-variance strategy as a purple line. It shows some similarities with the 60/40 strategy but performs worse than that one in the long run.

A final comparison is made with the strategy of investing exclusively in the risk-free security. We plot in Figure \ref{sec_emp_fig1} the wealth of our investor following such strategy as a green line. As expected, with the exception of the great depression, this strategy underperfoms all other strategies that we studied above. Clearly, it is not a recommended long term investment strategy.

So far we considered only the situation of an investor that set up her portfolio in 1925. However, it is well known that the performance of investment strategies depends on the starting date and the time horizon of the investor under consideration. Therefore, for completeness we do now investigate how our approach fares against the 60/40 strategy and the mean-variance strategy for different start dates and for different time horizons (equivalently the ending year of the respective portfolio strategy). Table \ref{table_prob2} presents the results for both strategies in separate panels. Note that by running through the respective diagonals, the reader can evaluate results for time horizons from 10 to 90 years in 10 year intervals. 
Table \ref{table_prob2} reinforces our earlier conclusions based on Figure \ref{sec_emp_fig1}. It shows that our strategy would have outperformed the 60/40 in the long run, over all horizons (at least 10 years long) and outperformed the mean-variance strategies over all time periods considered, with the exception of the time period 1965-1975.

\section{Conclusion}\label{sec_concl}

This paper has studied investing for the long run under expected utility maximization in a more general framework than the classical no-arbitrage theory provides. This has lead us introduce a stochastic discount factor (SDF) that is more general than the classical one. We have stressed the importance of its tradeability and identified it as the inverse of the Growth Optimal Portfolio (GP). Using this SDF we have proceeded analogous to the martingale technique and characterized the optimal consumption-savings and investment strategy via the GP. This has turned our attention to stochastic process properties of the GP. We showed that multiple fund separation theorems hold. Therein, the fund holdings are characterized via partial derivatives of the investor's value function, as long as the GP is part of a Markovian state process. Finally, we evaluated our strategy empirically.

\newpage 

\section*{Appendix: Additional Material}

\renewcommand{\thesection}{\Alph{section}}
\renewcommand{\thesubsection}{\Alph{section}.\arabic{subsection}}
\renewcommand{\theequation}{\Alph{section}-\arabic{equation}}
\renewcommand{\thetheorem}{\Alph{section}-\arabic{theorem}}
\setcounter{section}{0}  % reset counter 

\setcounter{theorem}{0}  % reset counter 
\setcounter{equation}{0}  % reset counter 

\section{Recursive Epstein-Zin Preferences}\label{appA}

Our setup allows us also to study consumption-savings problems ($\chi=1$) with preference structures that are more general than time-separable preferences, so-called recursive preferences. The function $f$ is called the (normalized) aggregator of current consumption and continuation utility.
A popular form of recursive preferences are the so-called Epstein-Zin preferences, introduced by \citet{Eps&Zin:89} based on the Kreps-Porteus preference specification. We now discuss these in detail.

\subsection{The Aggregator}\label{subsec_pref2}

To describe Epstein-Zin preferences we introduce the, so-called, elasticity of intertemporal substitution parameter $\psi>0$, the rate of time-preference  $\delta>0$, as well as, a risk-aversion coefficient $\gamma$, $0 < \gamma, \gamma \neq 1$. We then define a time-independent function $f$ for strictly positive $c>0$ and for strictly positive $l>0$ (strictly negative $l<0$) when $\gamma<1$ (when $\gamma>1$):
\begin{eqnarray}
f(c,l) & = & \delta \frac{1-\gamma}{1-\frac{1}{\psi}}  l \left( \left( \frac{c}{((1-\gamma) l )^{\frac{1}{1-\gamma}}} \right)^{1-\frac{1}{\psi}} -1 \right)  \text{ for } \psi \neq 1, \label{eq_EZ1} \\
f(c,l) & = & \delta (1-\gamma) l \left( \ln (c) - \frac{1}{1-\gamma} \ln \left( (1-\gamma) l \right) \right) \text{ for } \psi =1 . \label{eq_EZ2}
\end{eqnarray}

Leaving aside the multiplicative term in $\varepsilon$, our optimization problem yields then the \citet{Duf&Eps:92} parametrization of a price taking agent with stochastic differential utility derived from lifetime consumption. This characterizes also a continuous-time version of the \citet{Eps&Zin:89} preferences that permit separation of risk aversion from the intertemporal rate of substitution. Throughout the current paper, we do allow explicitly for a bequest function $\varepsilon B(V_T^\pi)$. Similarly to \citet{Liu:98}, the parameter $\varepsilon > 0$ allows us to adjust the relative importance of bequest and lifetime consumption.

Setting $\psi = 1/\gamma$ in equation \eqref{eq_EZ1} reduces the above recursive utility consumption-savings problem to a consumption savings problem with time-separable CRRA preferences and (relative) risk aversion coefficient $\gamma$. This restriction has been imposed in the literature to compare results with those of the, so-called, Merton consumption-savings problem, see \citet{Mer:71}. 

To see that the aggregator function $f$ in equations (\ref{eq_EZ1}) and (\ref{eq_EZ2}) fulfills the conditions in Assumption \ref{ass2} we note first that the function $f$ is twice differentiable on $\mathbb{R}^+$. Taking derivatives based on equation \eqref{eq_EZ1} ($\psi \neq 1$) and based on equation \eqref{eq_EZ2} ($\psi = 1$) we obtain:
\begin{eqnarray}
\frac{\partial f}{\partial c}  & = & 
%\delta (1-\gamma) l \frac{c^{-\frac{1}{\psi}} }{((1-\gamma) l )^{\frac{1-\frac{1}{\psi}}{1-\gamma}}} = 
\delta \left((1-\gamma) l \right)^{\frac{\frac{1}{\psi}-\gamma}{1-\gamma}} c^{-\frac{1}{\psi}} 
>0 , 
\text{ and } \
\frac{\partial^2 f}{\partial c^2} = -\frac{1}{\psi c} \frac{\partial f}{\partial c} <0 . \label{eq_EZ_deriv2} \label{eq_EZ_deriv3} \label{eq_EZ_deriv1}
\end{eqnarray}
Based on this representation it is straightforward to check the Inada conditions. 
We note that \citet{Duf&Lyo:92} and \citet{Scr&Ski:99} provide conditions and proofs for existence and uniqueness of lifetime utility $J$.

\subsection{Optimal Wealth and Consumption-savings Decision}

For further illustration throughout this subsection we use the bequest function $B(x)=e^{-\delta T} x^{1-\gamma}/(1-\gamma)$ for $\gamma \neq 1$ and $B(x)=e^{-\delta T} \ln (x)$ for $\gamma=1$. 
Equations (\ref{eq_EZ_deriv1}) provide the first-order derivatives of the aggregator $f$ that then allows us to derive the inverse w.r.t. consumption. We have 
$f^{\prime,-1}(x,l,s)=\delta^\psi \left( (1-\gamma) l \right)^{\frac{1-\gamma \psi}{1-\gamma}} x^{-\psi}$
and $B^{\prime,-1}(x)=e^{-\delta T} /x$. This gives for $0 \leq s < T$ the optimal consumption
\begin{eqnarray}
C_s^* & = & \delta^\psi \left( (1-\gamma) J_s \right)^{\frac{1-\gamma \psi}{1-\gamma}} \lambda^{-\psi} (D_s V_s^{GP})^\psi ,  \label{rec_pref_sol1}
\end{eqnarray}
and terminal value
\begin{eqnarray}
V_T^* = e^{-\delta T} \left( \frac{\varepsilon }{\lambda} V_T^{GP} \right)^{1/\gamma} .   \label{rec_pref_sol2}
\end{eqnarray}
It allows us to calculate the optimal benchmarked portfolio value function defined in equation \eqref{eq_defn_Uhat} as:
\begin{eqnarray}
&& \frac{V^* (t,v)}{v} \label{eq_optV_RecUt} \\
& = & E \left[ \left. \int_t^T \left( (1-\gamma) J_s \right)^{\frac{1-\gamma \psi}{1-\gamma}} \left(\frac{\delta D_s}{\lambda} \right)^\psi  (V_s^{GP})^{\psi-1} ds + e^{-\delta T} \left(\frac{\varepsilon }{\lambda} \right)^{\frac{1}{\gamma}} (V_T^{GP})^{\frac{1}{\gamma}-1} \right| 
V_t^{GP} =v
\right]   .  \nonumber
\end{eqnarray}

The Lagrange multiplier $\lambda$ follows from solving the initial budget equation $V_0=V_0^*=V^* (0,F_0)$. This provides a full characterization of the optimal value process.

If we assume the intertemporal elasticity of substitution as $\psi=1/\gamma$, then we have $1-\gamma \psi=0$ such that 
$\left( (1-\gamma) J_s \right)^{\frac{1-\gamma \psi}{1-\gamma}}=1$ and the process $J$ does no longer play a role in equation \eqref{eq_optV_RecUt}. We are then back in the representation presented in the previous subsection with the value function \eqref{eq_CRRA_Vstar}. This is as expected, since it is well-known that the case $\psi=1/\gamma$ corresponds to time-additive CRRA preferences.

The case without bequest ($\varepsilon=0$) allows us to further simplify this result. In that case we can factor out $\delta^\psi \lambda^{-\psi} (1-\gamma)^{(1-\gamma \psi)/(1-\gamma)}$ and using $V_0=V^*(0,V_0^{GP})$ we find based on \eqref{eq_optV_RecUt} that at all times $0 \leq t \leq T$ one has
\begin{equation}
V^*(t,v) = V_0 v \frac{E\left[ \left. \int_t^T J_s^{\frac{1-\gamma \psi}{1-\gamma}} D_s^\psi (V_s^{GP})^{\psi-1} ds \right| V_t^{GP} =v \right]}{E\left[ \int_0^T J_s^{\frac{1-\gamma \psi}{1-\gamma}} D_s^\psi (V_s^{GP})^{\psi-1} \right]} \ .  \label{eq_rec_Vstar}
\end{equation}
If we further assume the intertemporal elasticity of substitution as $\psi=1$, then the function $V^*$ simplifies to
\begin{eqnarray*}
V^* (t,v) & = & V_0 v \frac{E\left[ \left. \int_t^T J_s D_s ds \right| V_t^{GP} =v\right]}{E\left[ \int_0^T J_s D_s \right]} .
\end{eqnarray*}
Further analysis of this equation and of \eqref{eq_rec_Vstar} requires among others studying the distributional properties of the GP.

%\subsection{Portfolio Choice with Long-Run Risks}
%
%\citet{Ban&Yar:04} introduced the lon-run risk (LRR) model to study the price impact of two forms of risk: a small long-run predictable component driving consumption and fluctuating economic uncertainty measures by consumption volatility. While their model is set up in discrete time, we are interested in the continuous-time analogue characterized by 
%\begin{eqnarray*}
%d \ln C_t & = & (\mu + x_t) dt + \sigma_t dW_t^1 \\
%d \ln D_t & = & (\mu_d + \phi x_t) + \phi_d \sigma_t dW_t^2 \\
%d x_t & = & (\rho-1) x_t dt + \sigma_t dW_t^3 \\
%d \sigma_t^2 & = & (1-\nu_1) (\bar{\sigma}^2 -\sigma_t^2) dt + \sigma_w d W_t^4.
%\end{eqnarray*}
%Here $(C_t)_t,(D_t)_t$ denote the (exogenously specified) processes of aggregate consumption and aggregate dividends and the parameters $\mu,\mu_d, \phi, \phi_d, \nu_1, \sigma_w$ are given. At times, it is set $W^1=W^2$. 
%
%I THINK IT WOULD BE NICE TO SAY SOMETHING ABOUT INVESTING FOR A SMALL INVESTOR (WITH RECURSIVE PREFERENCE) FACING A PRICE-SETTING REPRESENTATIVE INVESTOR AS IN THAT LITERATURE. (SO FAR OPTIMAL INVESTING HAS NOT BEEN STUDIED.) 

\section{Optimal Portfolio Value for Particular GP Processes}

Section \ref{sec3} identifies properties of the GP that would allow us to find the minimal possible price for a targeted payoff, while subsequent sections show how to hedge this payoff conveniently by only investing in a proxy of the GP and the baseline security. This lead us in Section \ref{sec_attain} to study tradeable proxies of the GP. 

In addition, from a modeling perspective, this draws our attention to properties of the GP and modeling its dynamics adequately. Constructing a proxy of the GP and interpreting it as GP would then avoid the practically challenging and almost impossible task of modeling and estimating all those factors and parameters that determine the entire market dynamics and that, in principle, one needs to have access to in order to identify accurately the theoretically precise GP. Therefore, we will assume in the following that we have constructed a tradeable proxy of the GP and, therefore focus on exploring properties of the GP that are sufficient for hedging in consumption-savings investments.

It is important to recall that different to the well-known martingale technique, see \citet{Pen:08}, and \citet{Cvi&Zap:04}, we do not assume the existence of an equivalent risk neutral probability measure. This allows us to use a much richer modeling world when characterizing the dynamics of the market. The first subsection below assumes a classical market model with a constant investment opportunity, whereas the second subsection considers a market model that is more realistic and does not fit any longer into the world of classical market models.

\subsection{Constant Investment Opportunity Set}

A most convenient case has been widely studied in the literature, where one assumes a Black-Scholes dynamics with constant volatility for the GP. When we assume a constant market price of risk $\theta_t=\theta>0$, we obtain for any $0 \leq t \leq s \leq T$ for the discounted GP the expression
\begin{eqnarray*}
V_s^{GP} & = & V_t^{GP} \exp\left\{ \frac{\theta^2}{2} (s-t) + \theta (W_s -W_t) \right\}    \\
\text{ and so } E \left[ \left. \left(V_s^{GP}\right)^{(1/\gamma)-1} \right| V_t^{GP} \right] & = & \left( V_t^{GP} \right)^{(1/\gamma)-1} \exp \left\{ \theta^2 \frac{ 1-\gamma}{2 \gamma^2} (s-t) \right\} .
\end{eqnarray*}
This allows us to express the value function $V^*$ in terms of the GP as in Subsection \ref{subsec_scalar_GP}. We then derive in the case of preferences over terminal wealth based on \eqref{eq_TW_value} the benchmarked value function 
\begin{equation}
\hat{V}_t^* = \frac{V^* (t,V_t^{GP})}{V_t^{GP}} = V_0 \exp \left( \theta^2 \frac{\gamma -1}{2 \gamma^2} t \right) \left( V_t^{GP} \right)^{(1/\gamma)-1}  \ .
\end{equation}

Multiplying this expectation through with $V_t^{GP}=v$, yields $V^* (t, V_t^{GP}$. By taking the first order derivative of $V^* (t,V_t^{GP})$ w.r.t. $v$, we note that the right hand side in this equation is proportional to $V^*(t,V_t^{GP}$, that is
$$
\frac{\partial V^* (t,V_t^{GP})}{\partial V_t^{GP}} = \frac{1}{\gamma} \frac{V^* (t,V_t^{GP})}{v} = \frac{1}{\gamma} \hat{V}_t^* .
$$
According to Theorem \ref{tfsep_thm}, therefore, $V^* (t,V_t^{GP})$ also denotes $\gamma$ times the number of units of the GP the investor holds.

Next, we look at the case of time-additive CRRA preferences, where we write
$$
\rho = \theta^2 \frac{ 1-\gamma }{2 \gamma^2} -\delta/\gamma .
$$
We derive that
\begin{eqnarray*}
& & E \left[ \left. \int_t^T e^{-(\delta/\gamma) s} (V_s^{GP})^{1/\gamma-1} ds + \varepsilon e^{-(\delta/\gamma) T} (V_T^{GP})^{1/\gamma-1} \right |V_t^{GP} =v \right] \\
& = & \left( V_t^{GP} \right)^{1/\gamma-1} e^{-(\delta/\gamma) t} \int_t^T \exp \left\{ \rho (s-t) \right\} ds + \varepsilon \left( V_t^{GP} \right)^{1/\gamma-1} e^{-(\delta/\gamma) t} \exp \left\{ \rho (T-t) \right\} \\
& = & \left( V_t^{GP} \right)^{1/\gamma-1} e^{-(\delta/\gamma) t} \frac{1}{\rho } \left( \exp \left\{\rho (T-t) \right\} -1\right) + \varepsilon \left( V_t^{GP} \right)^{1/\gamma-1} e^{-(\delta/\gamma) t} \exp \left\{ \rho (T-t) \right\}
\end{eqnarray*}

In the case of time-additive CRRA preferences we then find based on \eqref{eq_CRRA_Vstar} that
\begin{eqnarray*}
\frac{V^* (t,v)}{v}
& = & V_0 v^{1/\gamma-1} e^{-(\delta/\gamma) t}
 \cdot   \frac{\frac{1}{\rho} \left( \exp \left\{ \rho (T-t) \right\} -1\right)
+
\varepsilon \exp \left\{ \rho (T-t) \right\}
}{
\frac{1}{\rho} \left( \exp \left\{  \rho T \right\} -1\right)
+
\varepsilon \exp \left\{ \rho T \right\}
} \ .
\end{eqnarray*}

\subsection{Minimal Market Model}\label{app_MMM}

In the Minimal Market Model (MMM) of \citet{Pla:01}, see also Chapter 13 in \citet{Pla&Hea:10}, the (discounted) GP $V_t^{GP}$ can be expressed as the product 
\begin{equation}
V_t^{GP} = Y_t \alpha_t
\end{equation}
of a square root process $(Y_s)_{0 \leq s \leq T}$ of dimension four with an exponential function of time $\alpha_t = \alpha_0 \exp( \eta t), \eta >0, \alpha_0>0$, where
\begin{equation}
d Y_t = (1 - Y_t) \eta dt + \sqrt{\eta Y_t} d \tilde{W}_t
\end{equation}
for $0 \leq t < \infty, Y_0 =\frac{1}{\alpha_0}$, with $\tilde{W}$ denoting a Brownian motion. The volatility of $V_t^{GP}$ equals then that of $Y_t$, which is
\begin{equation}
\tilde{b}_t^F = \sqrt{\frac{\kappa}{Y_t}} = \sqrt{\frac{\kappa \alpha_t}{V_t^{GP}}} .
\end{equation}

The only two parameters needed are $\alpha_0>0$ and $\eta$. Both can be fitted by noting that the quadratic variation of $\sqrt{V_t^{GP}}$ equals
\begin{equation}
\varphi(t)= <\sqrt{V_t^{GP}}>_t = \frac{\alpha}{4} \left( e^{\eta t} -1 \right), 
\end{equation}
such that
\begin{equation}
\eta = \frac{1}{t} \ln \left( \frac{<\sqrt{V_t^{GP}}>_t}{\alpha_0} +1 \right)  , 
\end{equation}
which allows one to estimate $\eta$ and also $\alpha_0$ for sufficiently long time periods.

As shown in \citet{Pla&Hea:10}, the MMM does not fit under the classical no-arbitrage paradigm because the inverse of the discounted GP is not a true martingale and only a strict local martingale. Therefore, the density of the putative risk-neutral measure $\frac{V_0^{GP}}{V_t^{GP}}$ is only a strict local martingale, and an equivalent risk-neutral probability measure does not exist. From equation (8.7.14) in \citet{Pla&Hea:10} it follows for $\gamma \in (1, \infty]$ and $0 \leq t \leq s < \infty$ that
\begin{eqnarray}
&& E[(V_s^{GP})^{(1/\gamma)-1}|V_t^{GP}] \\
& = & (2 (\varphi(s)- \varphi(t))^{(1/\gamma)-1} \exp \left\{ - \frac{V_t^{GP}}{2 (\varphi(s) - \varphi(t))} \right\} \sum_{k=0}^\infty \left( \frac{V_t^{GP}}{2 (\varphi(s) -\varphi(t))} \right)^k \frac{\Gamma(\frac{1}{\gamma}+1+k)}{k! \ \Gamma (k+2)} , \nonumber 
\end{eqnarray}
where
\begin{eqnarray}
\varphi (t) & = & \frac{\alpha_0}{4 \eta} \left( e^{ \eta t } -1 \right) ,
\end{eqnarray}
and $\Gamma$ denotes the Gamma function. For instance, in the special case $\gamma=1/2$ we have
\begin{equation}
E[V_s^{GP} |V_t^{GP}] = 4 (\varphi(s) -\varphi(t)) + V_t^{GP} .
\end{equation}

In the case $\gamma=1$ we have log-utility and obtain 
$$
E[(V_s^{GP})^0| V_t^{GP}] = 1
$$
which yields an important simplification of any calculation.

%In the case of extreme risk aversion $\gamma=\infty$ we get
%$$
%E[(V_s^{GP})^{-1} |V_t^{GP}] = V_t^{GP} ( 1 - \exp \{ \ldots \}) .
%$$
%
%This allows us to calculate $V^*(t,v)$ according to \eqref{eq_CRRA_Vstar} as
%$$
%V^* (t,v) = ...
%$$

\bibliography{biblio}

\end{document}